\documentclass[12pt,prd,preprint,nofootinbib]{revtex4-1}

\usepackage{latexsym,amsmath,amssymb,lmodern,float,url}
\usepackage[table]{xcolor}
\usepackage{array}
\usepackage{booktabs}
\usepackage{bm} 
\usepackage{graphicx,epsfig}
\usepackage{color}
\usepackage{slashed}
\usepackage{longtable}
\usepackage{braket}
\usepackage{pifont}
\usepackage[pdftex,bookmarks,linktocpage,plainpages=false,hyperfigures,linkcolor=blue,citecolor=blue,urlcolor=blue]{hyperref} 
\hypersetup{colorlinks=true}

\def\Dslash{D\!\!\!\!/\,}
\def\qbar{\bar{q}}
\def\Nbar{\bar{N}}

\newcommand\ee{\end{equation}}
\newcommand\be{\begin{equation}}
\newcommand\eea{\end{eqnarray}}
\newcommand\bea{\begin{eqnarray}}
\newcommand\aea{&=&}

\newcommand\R{\cellcolor{red!40}\ding{51}}
\newcommand\Red{\cellcolor{red!80}\ding{51}}
\newcommand\G{\cellcolor{green!68}\ding{51}}
\newcommand\B{\cellcolor{blue!38}\ding{51}}
\newcommand\rB{\cellcolor{blue!98}\ding{51}}
\newcommand\Or{\cellcolor{orange!68}\ding{51}}
\newcommand\Pu{\cellcolor{purple!78}\ding{51}}
\newcommand\Y{\cellcolor{yellow!78}\ding{51}}
\newcommand\g{\cellcolor{gray!58}\ding{51}}
\newcommand\C{\cellcolor{cyan!58}\ding{51}}
\newcommand\M{\cellcolor{magenta!68}\ding{51}}
\newcommand\x{\ding{51}}

\newcommand\half{\frac{1}{2}}

\newcolumntype{C}[1]{>{\centering\let\newline\\\arraybackslash\hspace{-2pt}}m{#1}}

\newcommand\bfootnote[1]{%
  \begingroup
  \renewcommand\thefootnote{}\footnote{#1}%
  \addtocounter{footnote}{-1}%
  \endgroup
}
\begin{document}

\title{A General Analysis of Direct Dark Matter Detection:  From Microphysics to Observational Signatures}

\author{James B. Dent$^{\bf a}$}
\author{Lawrence M. Krauss$^{\bf b, c}$}
\author{Jayden L. Newstead$^{\bf b}$}
\author{Subir Sabharwal$^{\bf b}$}

\affiliation{$^{\bf a}$ Department of Physics, University of Louisiana at Lafayette, Lafayette, LA 70504, USA,}

\affiliation{$^{\bf b}$ Department of Physics and School of Earth and Space Exploration, Arizona State University, Tempe, AZ 85287, USA,}

\affiliation{$^{\bf c}$ Research School of Astronomy and Astrophysics, Mt. Stromlo Observatory, Australian National University, Canberra 2614, Australia}

\date{May 12 2015}

\begin{abstract}

Beginning with a set of simplified models for spin-0, spin-$\half$, and spin-1 dark matter candidates using completely general Lorentz invariant and renormalizable Lagrangians, we derive the full set of non-relativistic operators and nuclear matrix elements relevant for direct detection of dark matter, and use these to calculate rates and recoil spectra for scattering on various target nuclei. This allows us to explore what high energy physics constraints might be obtainable from direct detection experiments, what degeneracies exist, which operators are ubiquitous and which are unlikely or sub-dominant. We find that there are operators which are common to all spins as well operators which are unique to spin-$\half$ and spin-1 and elucidate two new operators which have not been previously considered. In addition we demonstrate how recoil energy spectra can distinguish fundamental microphysics if multiple target nuclei are used.  Our work provides a complete roadmap for taking generic fundamental dark matter theories and calculating rates in direct detection experiments. This provides a useful guide for experimentalists designing experiments and theorists developing new dark matter models.

\end{abstract}

\maketitle

\section{Introduction}

The existence of non-baryonic dark matter has been inferred from measurements including galactic rotation curves \cite{Sofue:2000jx}, large scale structure surveys \cite{Cole:2005sx,Beutler:2011hx,Anderson:2012sa}, X-ray observations \cite{Vikhlinin:2008ym}, gravitational lensing \cite{Fu:2007qq,Massey:2007gh}, and cosmic microwave background anisotropy measurements \cite{Ade:2013zuv}, spanning cosmological eras from the present day to the remote past.  This widespread and robust data has led to cold dark matter models with a cosmological constant, labeled  $\Lambda$CDM  becoming entrenched as the standard cosmological model.

Nevertheless, this impressive array of observations has only been sensitive to the \emph{gravitational} influence of dark matter and constrained its relic abundance, leaving its particle nature as one of the most important open questions in physics.  The search for dark matter includes indirect astrophysical searches (\cite{Adriani:2008zr,Su:2010qj,Hooper:2010mq,Weniger:2012tx,Bulbul:2014sua}), collider production efforts (for some examples of dark matter searches at the LHC, see \cite{Aad:2014vka,Aad:2014vea,Aad:2015zva,Aad:2013oja,Lowette:2014yta}) which will examine new territory soon with LHC run 2 which will commence this year, and attempts to observe dark matter interactions with Standard Model (SM) particles via dark matter-nucleus scattering processes in direct detection experiments, to which we now turn.

The search for dark matter via direct detection goes back at least three decades \cite{Goodman:1984dc,Cabrera:1984rr} and has been particularly vigorous over the last decade or so with experiments such as LUX \cite{Akerib:2013tjd}, Xenon100 \cite{Aprile:2012nq}, CDMS II (Ge) \cite{Ahmed:2010wy}, CDMS I (Si) \cite{Agnese:2013rvf}, DAMA/LIBRA \cite{Bernabei:2008yi}, COGENT \cite{Aalseth:2012if}, and CRESST \cite{Angloher:2011uu}  pushing ever deeper into weakly interacting dark matter mass and scattering cross-section parameter space, but has thus far failed to yield a convincing signal.  In the near future detectors such as Super CDMS \cite{Sander:2012nia} (which has recently released its first results on low mass dark matter searches \cite{Agnese:2014aze,Agnese:2013jaa}), XENON1T \cite{Aprile:2012zx}, and DARWIN \cite{Baudis:2012bc} are expected to push the limits of direct detection orders of magnitude below the current levels.

In order to connect observations to microphysical models one needs a general framework within which to interpret the observations of direct detection experiments.  For quite some time the prevailing method of analyzing dark matter-nucleus interactions has been to assume that dark matter is a weakly interacting massive particle (WIMP), and then to categorize the interactions as  elastic and isospin conserving and either spin-independent or spin-dependent \cite{Pato:2010zk,Newstead:2013pea}.  For some well studied models of dark matter, such as the weakly interacting Majorana neutralino found in supersymmetry models, this assumption is reasonable.  

With an absence of observed dark matter signals, there has of late been a surge in interest in exploring more general types of interactions between dark matter and nuclei.  Generalizations include inelastic and momentum dependent interactions, which may arise due to additional structure in the dark sector including excited dark matter states, or dark gauge bosons giving rise to electric and magnetic form factors~\cite{TuckerSmith:2001hy,Chang:2008gd,Feldstein:2009tr,Chang:2009yt,Barger:2010gv,Banks:2010eh,Foot:2010hu,Ho:2012bg}. 

The formalism of choice for many of these investigations is relativistic effective field theory, which provides a model independent framework to analyse dark matter-SM interactions~\cite{Goodman:2010ku,Bai:2010hh,Balazs:2014rsa}. It has been shown that these effective theories break down when applied to high-momentum transfer experiments, such as the LHC~\cite{Busoni:2013lha}. Therefore analyses moved beyond this framework and have moved to what are labeled as `simplified models' instead~\cite{Alves:2011wf,Rajaraman:2011wf,Buchmueller:2013dya}. Simplified models are field theories which extend the SM by a single dark matter particle and a single mediator particle which allows the WIMP to communicate with quarks and/or leptons. The newly added dark matter content is assumed to be a singlet under the SM gauge groups (we will consider some cases where the particles mediating the interaction have SM charge). In this context it is then possible to calculate collider amplitudes valid at the high energies of interest in such experiments. Given this simple dark sector, one can write down an exhaustive list of every combination of WIMP and mediator spins, and all possible tree level interactions. These simplified models have now gained popularity for analyzing indirect detection signals~\cite{Berlin:2014tja,Balazs:2014jla}, allowing connections to be made with the growing body of literature which make use of them. \\

Another step towards placing dark matter-nucleus interactions on a general footing has been accomplished recently by utilizing a non-relativistic effective field theory (EFT) approach \cite{Fan:2010gt,Fitzpatrick:2012ix,Anand:2013yka,Anand:2014kea}.  Since the interactions in direct detection scenarios are assumed to take place due to an incoming dark matter particle with a typical velocity $\mathcal{O}(100$km/s), the recoil momenta in such an interaction will be $\mathcal{O}(\lesssim 100$keV).  The particle masses involved, including the nucleons of roughly GeV scale, the dark matter particles, which typically range from the GeV region to several orders of magnitude above, and mediators that can also be quite heavy compared to the typical interaction momenta, produce a situation where an EFT treatment is quite natural. \\
  
In order to circumvent as much model dependence as possible, one can construct general interactions which obey Galilean invariance, $T$-symmetry, and Hermiticity.  These operators will take the standard effective four-particle interaction form, reminiscent of Fermi's original model of weak interactions.   The non-relativistic interactions can be shown to be functions of only four parameters including the nucleon spin $S_N$, the dark matter spin $S_{\chi}$, the momentum transfer, $\vec{q}$, and a kinematic variable $\vec{v}^{\perp}$ which is a function of the relative incoming ($\vec{v}_{\chi,in} - \vec{v}_{N,in}$) and outgoing velocities $\vec{v}_{\chi,out} - \vec{v}_{N,out}$ 
\bea
\vec{v}^{\perp} = \frac{1}{2}\left(\vec{v}_{\chi, in} - \vec{v}_{N, in} + \vec{v}_{\chi, out} - \vec{v}_{N, out}\right) = \vec{v}_{\chi, in} - \vec{v}_{N, in} + \frac{\vec{q}}{2\mu_N}
\eea
which obeys $\vec{v}^{\perp}\cdot \vec{q} = 0$.  It was demonstrated in \cite{Fitzpatrick:2012ix} that there exist fifteen such non-relativistic interactions which arise from twenty possible bi-linear combinations of dark matter and nucleons.

The formalism developed in \cite{Fitzpatrick:2012ix} is unique in being the only analysis to comprehensively develop the nuclear physics of direct detection experiments.  From this general framework it is now apparent that there are interactions beyond the standard spin independent/dependent type.  The origins of these `new' interactions are not necessarily exotic and it has been shown, in the context of relativistic EFT, how many of them can be generated \cite{Gresham:2014vja}.

What has been lacking to date however, is a completely general and comprehensive treatment that connects high energy microphysics with low-energy effective nuclear matrix elements in a model independent way.  It is possible, for example, that the various interactions listed in \cite{Fitzpatrick:2012ix} can give rise to degeneracies where different fundamental dark matter Lagrangians, describing dark matter and interaction mediators of various spins, can produce the same interaction types.  This will obviously pose problems for attempts to discern the properties of dark matter when interpreting the results of experimental data.  Furthermore, dark matter may not be spin-$\half$, which creates a need for extending the parametric framework from the four descriptors listed above.  In particular, as we shall show, this allows the existence of new non-relativistic operators to appear in the low energy effective theory.   

Motivated by the above we present here a general analysis covering a broad spectrum of particle and interaction types, starting from the microphysics, which will enable one to link experiment with fundamental theory while incorporating the new nuclear responses described in~\cite{Fitzpatrick:2012ix}. 

In this work we build upon the NR-EFT description by examining simplified models which incorporate the most general renormalizable Lagrangians for scalar, spinor, and vector dark matter interacting with nucleons via scalar, spinor, and vector mediators, consistent with Lorentz invariance and hermiticity while imposing stability of the dark matter candidates. We integrate out the heavy mediator and obtain effective relativistic interaction Lagrangians. Next, we take the non-relativistic limit of these Lagrangians, and identify them with the NR operators from~\cite{Fitzpatrick:2012ix}, which are reproduced below, in Table $1$. Using these, we identify which electroweak nuclear responses are excited by a given fundamental interaction model and determine the relative importance of various models within the context of direct detection experiments consisting of xenon and germanium targets by exploring the relative magnitude of coefficients of these operators, and also their energy dependence.

The paper is organized as follows; in section~\ref{secEFT} the EFT formalism of~\cite{Fitzpatrick:2012ix} is summarized, in section~\ref{secSimple} we build the generalized relativistic Lagrangians and in section~\ref{secNRR} we outline the signatures and distinguishability of these models in the context of direct detection experiments, providing a framework for both experimentalists and theorists to base their future analyses.

\section{Effective field theory of direct detection}
\label{secEFT}
Conventionally, coherent WIMP-nucleus scattering has been considered to come from two types of interactions; spin-independent (SI) and spin-dependent (SD). SI interactions couple to the charge/mass of the nucleus while SD couples to the spin. The nuclear cross section is generally written in terms of the nucleon cross section at zero momentum transfer, $\sigma_0$, and a form factor, $F(q)$, to take into account the loss of coherence over the finite size of the nucleus,
\be
\frac{d\sigma}{dE_r} = \frac{M}{2 \pi \mu_{\chi M} v^2}\left(\sigma_0^{SI} F_{SI}^2(q) + \sigma_0^{SD} F_{SD}^2(q)\right).
\ee
where $M$ is the mass of the target nucleus and $\mu_{\chi M}$ is the WIMP-nucleus reduced mass. 
This picture has recently been shown to be incomplete, as it is also possible for the WIMP to couple to the nucleus through additional nuclear responses \cite{Fitzpatrick:2012ix}. Working in the language of a non-relativistic (NR) effective field theory Fitzpatrick et al.~identified 15 operators to characterize the ways in which a WIMP can couple to the various nuclear responses. These operators are constructed from combinations of non-relativistic vectors which respect Galilean invariance, $T$ symmetry and which are Hermitian.  We list them in table~\ref{tabHaxtonOp}. The Hermitian vectors are:
\bea
i\frac{\vec{q}}{m_N},\ \vec{v}^{\perp}=\vec{v}+\frac{\vec{q}}{2\mu_N}, \ \vec{S}_\chi, \ \vec{S}_N,
\eea
where $\vec{q}=\vec{p'}-\vec{p}=\vec{k}-\vec{k'}$ is the momentum transfer, $\vec{v}$ is the velocity of WIMP with respect to the nucleus of the detector, $\mu_N$ is the reduced mass of the system and $\vec{S}_\chi$ and $\vec{S}_N$ are the WIMP and nuclear spins respectively. Throughout the paper, we denote by $\vec{p}$ and $\vec{p'}$ the incoming and outgoing WIMP momenta and by $\vec{k}$ and $\vec{k'}$ the incoming and outgoing nuclear momenta respectively. Energy-momentum conservation implies the orthogonality condition $\vec{q}\cdot\vec{v}^\perp=0$.  Here we will briefly outline the procedure employed in \cite{Fitzpatrick:2012ix} in going from the NR operators to the final differential WIMP-nucleus cross section. \\

\begin{table}[ht]
\caption{List of NR effective operators described in~\cite{Fitzpatrick:2012ix}}
\begin{tabular}{rc}
$\mathcal{O}_1$ & $1_\chi 1_N$ \\
$\mathcal{O}_2$ & $(\vec{v}^{\perp})^2$ \\
$\mathcal{O}_3$ & $i\vec{S}_N\cdot(\frac{\vec{q}}{m_N}\times \vec{v}^{\perp})$ \\
$\mathcal{O}_4$ & $\vec{S}_\chi \cdot \vec{S}_N$ \\
$\mathcal{O}_5$ & $i\vec{S}_\chi\cdot(\frac{\vec{q}}{m_N}\times \vec{v}^{\perp})$ \\
$\mathcal{O}_6$ & $(\frac{\vec{q}}{m_N}\cdot\vec{S}_N)(\frac{\vec{q}}{m_N}\cdot\vec{S}_\chi)$ \\
$\mathcal{O}_7$ & $\vec{S}_N \cdot \vec{v}^{\perp}$ \\
$\mathcal{O}_8$ & $\vec{S}_\chi \cdot \vec{v}^{\perp}$ \\
$\mathcal{O}_9$ & $i\vec{S}_\chi \cdot (\vec{S}_N \times \frac{\vec{q}}{m_N}) $ \\
$\mathcal{O}_{10}$ & $i\frac{\vec{q}}{m_N} \cdot \vec{S}_N$ \\
$\mathcal{O}_{11}$ & $i\frac{\vec{q}}{m_N} \cdot \vec{S}_\chi$ \\
$\mathcal{O}_{12}$ & $\vec{S}_\chi \cdot(\vec{S}_N \times \vec{v}^{\perp})$ \\
$\mathcal{O}_{13}$ & $i(\vec{S}_\chi \cdot \vec{v}^{\perp})(\frac{\vec{q}}{m_N} \cdot \vec{S}_N)$ \\
$\mathcal{O}_{14}$ & $i(\vec{S}_N \cdot \vec{v}^{\perp})(\frac{\vec{q}}{m_N} \cdot \vec{S}_\chi)$ \\
$\mathcal{O}_{15}$ & $-(\vec{S}_\chi \cdot \frac{\vec{q}}{m_N})\left( (\vec{S}_N\times \vec{v}^{\perp})\cdot\frac{\vec{q}}{m_N}\right)$ \\
\end{tabular}
\label{tabHaxtonOp}
\end{table}

In general one can write down the non-relativistic interaction Lagrangian as
\bea
\mathcal{L}_{NR}=\sum_{\alpha=n,p}\sum_{i=1}^{15}c_i^\alpha\mathcal{O}_{i}^\alpha,
\eea
where the coefficients $c_{i}^\alpha$ are given by the microphysics of the interaction and in general one could allow for isospin violation by having different couplings to neutron and proton inside the nucleus. This can be rewritten in $2$-component isospin space as	
\bea
\mathcal{L}_{NR}=\sum_{\tau=0,1}\sum_{i=1}^{15}c_{i}^{\tau}\mathcal{O}_it^\tau
\eea
where $t^0$ and $t^1$ are the identity matrix and the Pauli matrix $\sigma^3$ respectively. The nucleus is composed of nucleons, and these can individually interact with the WIMP. This is incorporated by considering the operator $\mathcal{O}(j)$ as an interaction between a single nucleon, $j$, and the WIMP, and then summing over the nucleons. 
\bea
\label{eqnNRL}
\sum_{\tau=0,1}\sum_{i=1}^{15}c_{i}^{\tau}\mathcal{O}_it^\tau\rightarrow \sum_{\tau=0,1}\sum_{i=1}^{15}c_{i}^{\tau}\sum_{j=1}^A\mathcal{O}_i(j)t^\tau(j)
\eea
where $A$ is the atomic mass number given by the total number of neutrons and protons.  One can do the same reduction with $\vec{v}^{\perp}$,
\bea
\vec{v}^{\perp}&\rightarrow& \{\vec{v}_{\chi}-\vec{v}_N(i), i=1,...,A\}\nonumber\\
&\equiv& \vec{v}_{T}^{\perp}-\{\dot{\vec{v}}_N(i),i=1,...,A-1\}
\eea
where $\vec{v}_{\chi}$ and $\vec{v}_N(i)$ are the symmetrized combination of incoming and outgoing velocities for the WIMP and nucleons respectively. $\vec{v}_T^{\perp}$ (here $T$ stands for target, i.e., the nuclear center-of-mass) is defined as
\bea
\vec{v}_T^{\perp}=\vec{v}_{\chi}-\frac{1}{2A}\sum_{i=1}^A\left[\vec{v}_{N,in}(i)+\vec{v}_{N,out}(i)\right]
\eea
This allows for a decomposition of the nucleon velocities into internal velocities $\dot{\vec{v}}_N(i)$ that act only on intrinsic nuclear coordinates and `in' and `out' velocities that evolve as a WIMP scatters off the detector. As an example, the dot product between $\vec{v}_N^\perp$ and $\vec{S}_{N}$ can be rewritten as
\bea
\vec{v}^{\perp}\cdot\vec{S}_N&\rightarrow&\sum_{i=1}^A\frac{1}{2}\left[\vec{v}_{\chi,in}+\vec{v}_{\chi,out}-\vec{v}_{N,in}(i)-\vec{v}_{N,out}(i)\right]\cdot\vec{S}_{N}(i)\\
&=&\vec{v}^\perp_T\cdot\sum_{i=1}^A\vec{S}_{N}(i)-\left\{\sum_{i=1}^A\frac{1}{2}\left[\vec{v}_{N,in}(i)+\vec{v}_{N,out}(i)\right]\cdot\vec{S}_{N}(i)\right\}_{int}
\eea
The second term in the curly brackets is internal to the nucleus and acts as an operator on the `in' and `out' nucleon states. $\vec{v}_{N,in}$ can be replaced by $\vec{p}_{N,in}/M$ acting on the incoming state, which can in turn be replaced by $i\overleftarrow{\nabla}/M$, and similarly $\vec{p}_{N,out}/M$ by $-i\overrightarrow{\nabla}/M$ on the outgoing nuclear state. Finally, since the nucleus is non-zero in size and individual nucleons locally interact with the WIMP, nuclear operators built from $\mathcal{O}_i$ are accompanied by an additional spatial operator $e^{-i\vec{q}\cdot\vec{x}(i)}$ where $x(i)$ is the location of the $i^{th}$ nucleon inside the nucleus.\\

Starting from Eqn.~\ref{eqnNRL} and using the substitution rules for $\vec{v}^\perp$ and including a factor of $e^{-i\vec{q}\cdot\vec{x}_i}$, the interaction Lagrangian can be written as a sum of five distinct terms (nuclear electroweak operators) that only act on internal nucleon states. Their coefficients, on the other hand, act on WIMP `in' and `out' states. The WIMP-nucleus interaction can then be written as
\bea
\sum_{\tau=0,1}\left\{l_0^\tau S+l_{0}^{A\tau} T+\vec{l}^\tau_5\cdot\vec{P}+\vec{l}_M^\tau\cdot{Q}+\vec{l}_{E}^\tau\cdot\vec{R}\right\}t^\tau(i)
\eea
where
\bea
S&=&\sum_{i=1}^Ae^{-i\vec{q}\cdot\vec{x}_i}\nonumber\\
T&=&\sum_{i=1}^A\frac{1}{2M}\left[-\frac{1}{i}\overleftarrow{\nabla}_i\cdot\vec{\sigma}(i)e^{-i\vec{q}\cdot\vec{x}_i}+e^{-i\vec{q}\cdot\vec{x}_i}\vec{\sigma}(i)\cdot\frac{1}{i}\overrightarrow{\nabla}_i\right]\nonumber\\
\vec{P}&=&\sum_{i=1}^A\vec{\sigma}(i)e^{-i\vec{q}\cdot\vec{x}_i}\nonumber\\
\vec{Q}&=&\sum_{i=1}^A\frac{1}{2M}\left[-\frac{1}{i}\overleftarrow{\nabla}_ie^{-i\vec{q}\cdot\vec{x}_i}+e^{-i\vec{q}\cdot\vec{x}_i}\frac{1}{i}\overrightarrow{\nabla}_i\right]\nonumber\\
\vec{R}&=&\sum_{i=1}^A\frac{1}{2M}\left[\overleftarrow{\nabla}_i\times\vec{\sigma}(i)e^{-i\vec{q}\cdot\vec{x}_i}+e^{-i\vec{q}\cdot\vec{x}_i}\vec{\sigma}(i)\times\overrightarrow{\nabla}_i\right]
\eea
and
\bea\label{eqnells}
l_0^\tau&=&c_1^\tau+ic_5^\tau\vec{S}_\chi\cdot\left(\frac{\vec{q}}{m_N}\times\vec{v}^\perp_T\right)+c_8^\tau(\vec{S}_\chi\cdot\vec{v}_T^\perp)+ic_{11}^\tau\frac{\vec{q}\cdot\vec{S}_\chi}{m_N}\nonumber\\
l_0^{A\tau}&=&-\frac{1}{2}\left[c_7^\tau+ic_{14}^\tau\left(\vec{S}_\chi\cdot\frac{\vec{q}}{m_N}\right)\right]\nonumber\\
\vec{l}_5&=&\frac{1}{2}\left[c_3^\tau{i}\frac{\left({\vec{q}}\times\vec{v}^\perp_T\right)}{m_N}+c_4^\tau\vec{S}_{\chi}+c_6^\tau\frac{(\vec{q}\cdot\vec{S}_\chi)\vec{q}}{m_N^2}+c_7^\tau\vec{v}_T^\perp+ic_9^\tau\frac{(\vec{q}\times\vec{S}_\chi)}{m_N}+ic_{10}^\tau\frac{\vec{q}}{m_N}\right)\nonumber\\
&&\left. c_{12}^\tau(\vec{v}_T^\perp\times\vec{S}_{\chi})+ ic_{13}^\tau\frac{(S_{\chi}\cdot\vec{v}^{\perp}_T)\vec{q}}{m_N}+ic_{14}^\tau\left(\vec{S}_\chi\cdot\frac{\vec{q}}{m_N}\right)\vec{v}^\perp_{T}+c_{15}^\tau\frac{(\vec{q}\cdot\vec{S}_\chi)(\vec{q}\times\vec{v}^\perp_T)}{m_N^2}\right]\nonumber\\
\vec{l}_M&=&c_5^\tau\left(i\frac{\vec{q}}{m_N}\times\vec{S}_\chi\right)-\vec{S}_\chi c_8^\tau \nonumber\\
\vec{l}_E&=&\frac{1}{2}\left[c_3^\tau\frac{{\vec{q}}}{m_N}+ic^{\tau}_{12}\vec{S}_{\chi}-c_{13}^\tau\frac{(\vec{q}\times\vec{S}_\chi)}{m_N}-ic_{15}^\tau\frac{(\vec{q}\cdot\vec{S}_\chi)\vec{q}}{m_N^2}\right]\label{eqnElls}\\ \nonumber
\eea
The WIMP-nucleus amplitude, $\mathcal{M}$, can then be succinctly written as
\bea
\mathcal{M}=\sum_{\tau=0,1}\langle j_\chi,M_\chi;j_N,M_N|\left\{l_0^\tau S+l_{0}^{A\tau} T+\vec{l}^\tau_5\cdot\vec{P}+\vec{l}_M^\tau\cdot{Q}+\vec{l}_{E}^\tau\cdot\vec{R}\right\}t^\tau(i)|j_\chi,M_\chi;j_N,M_N\rangle.\nonumber\\
\label{eqnAmp}
\eea
By using spherical decomposition, the internal nuclear operators $S, T, P, Q$ and $R$ can be further rewritten in terms of standard nuclear electroweak responses as follows:
\bea
\mathcal{M}&=&\sum_{\tau=0,1}\langle j_\chi,M_{\chi f};j_N,M_{Nf}|\left(\sum_{J=0}\sqrt{4\pi(2J+1)}(-i)^J\left[l_0^\tau M_{J0;\tau}-il_{0}^{A\tau}\frac{q}{m_N}\tilde{\Omega}_{J0;\tau}(q)\right]\right.\\
& &\left.+\sum_{J=1}\sqrt{2\pi(2J+1)}(-i)^J\sum_{\lambda\pm1}(-1)^\lambda\left\{l_{5\lambda}^\tau[\lambda\Sigma_{J-\lambda;\tau}(q)+i\Sigma^{'}_{J-\lambda;\tau}(q)]\right.\right.\nonumber\\
& &\left.\left.-i\frac{q}{m_N}l^\tau_{M\lambda}[\lambda\Delta_{J-\lambda;\tau}(q)]-i\frac{q}{m_N}l^\tau_{E\lambda}[\lambda\tilde{\Phi}_{J-\lambda;\tau}(q)+i\tilde{\Phi}^{'}_{J-\lambda;\tau}(q)]\right\}\right.\nonumber\\
& &\left.+\sum_{J=0}^\infty\sqrt{4\pi(2J+1)}(-i)^J\left[il^{\tau}_{50}\Sigma^{''}_{J0;\tau}(q)+\frac{q}{m_N}l_{M0}^{\tau}\tilde{\Delta}^{''}_{J0;\tau}(q)+\frac{q}{m_N}l_{E0}^{\tau}\tilde{\Phi}^{''}_{J0;\tau}(q)\right]\right)|j_\chi,M_{\chi i};j_N,M_{Ni}\rangle\nonumber
\eea
Where there is an implicit sum over the nucleons,
\bea
\mathcal{O}_{JM;\tau}(q)\equiv\sum_{i=1}^A\mathcal{O}_{JM}(q\vec{x}_i)t^\tau(i),
\eea
and the various electroweak responses are defined as
\bea
M_{JM}(q\vec{x})&\equiv& j_J(qx)Y_{JM}(\Omega_x)\nonumber\\
\vec{M}^M_{JL}&\equiv& j_J(qx)\vec{Y}_{JLM}(\Omega_x)\nonumber\\
\Delta_{JM}&\equiv&\vec{M}^M_{JJ}(qx_i)\cdot\frac{1}{q}\vec{\nabla}_i\nonumber\\
\Sigma^{'}_{JM}&\equiv& -i\left\{\frac{1}{q}\vec{\nabla}_i\times\vec{M}^M_{JJ}(q\vec{x}_i)\right\}\cdot\vec{\sigma}(i)\nonumber\\
\Sigma^{''}_{JM}&\equiv& \left\{\frac{1}{q}\vec{\nabla}_i{M}_{JM}(q\vec{x}_i)\right\}\cdot\vec{\sigma}(i)\nonumber\\
\tilde{\Phi}^{'}_{JM}&\equiv& \left[\frac{1}{q}\vec{\nabla}_i\times\vec{M}^M_{JJ}(q\vec{x}_i)\right]\cdot\left[\vec{\sigma}(i)\times\frac{1}{q}\vec{\nabla}_i\right]+\frac{1}{2}\vec{M}^M_{JJ}(q\vec{x}_i)\cdot\vec{\sigma}(i)\nonumber\\
\Phi^{''}_{JM}&\equiv& i\left[\frac{1}{q}\vec{\nabla}_i{M}_{JM}(q\vec{x}_i)\right]\cdot\left[\vec{\sigma}(i)\times\frac{1}{q}\vec{\nabla}_i\right]\nonumber\\
\Sigma_{JM}&\equiv&\vec{M}_{JJ}^M(q\vec{x}_i)\cdot\vec{\sigma}(i)\nonumber\\
\tilde{\Omega}_{JM}&\equiv&\Omega_{JM}(q\vec{x}_i)+\frac{1}{2}\Sigma^{''}_{JM}(q\vec{x}_i)\nonumber\\
\tilde{\Phi}_{JM}&\equiv&\Phi_{JM}(qx_i)-\frac{1}{2}\Sigma^{'}_{JM}(qx_i)\nonumber\\
\tilde{\Delta}^{''}_{JM}&\equiv&\Delta^{''}_{JM}(qx_i)-\frac{1}{2}M_{JM}(qx_i)\\ \nonumber
\eea
where $Y_{JM}$ and $\vec{Y}_{JLM}$ are spherical harmonics and vector spherical harmonics respectively. We are only considering elastic transitions, and assuming parity and CP as symmetries of the nuclear ground state. This eliminates some of the responses, and only $M, \Phi^{''}, \Sigma^{'}, \Delta, \Sigma^{''}, \tilde{\Phi}^{'}$ survive. To calculate cross-sections, one needs to square the amplitude, average over initial spins and sum over final spins. The matrix element squared for the nuclear portion of the amplitude has been made available by Fitzpatrick et al.~\cite{Fitzpatrick:2012ix}, and codes have been supplied to calculate the full amplitude and rate~\cite{Anand:2013yka}. 

As we shall describe, in the following analysis we discovered that two additional NR operators are required to fully describe the scattering of spin-1 WIMPs off nuclei, 
\bea
\mathcal{O}_{17}\equiv i\frac{\vec{q}}{m_N}\cdot\mathcal{S}\cdot\vec{v}_{\perp},\nonumber\\ \mathcal{O}_{18}\equiv i\frac{\vec{q}}{m_N}\cdot\mathcal{S}\cdot\vec{S}_{N}, \\ \nonumber
\eea
where $\mathcal{S}$ is the symmetric combination of polarization vectors. Appendix~\ref{appVec} contains the details required to include these new operators in the above formalism.

\section{Simplified Models for Direct Detection}
\label{secSimple}
From a model building perspective, one would like to know how relevant the novel nuclear responses are in interpreting direct detection data. Previous work \cite{Gresham:2014vja} demonstrated that using only the SI/SD form factors (even with additional momentum dependence taken into account) can lead one to infer wildly incorrect values of the WIMP mass and cross sections. 

Here we go further by starting with simplified models at the Lagrangian level, where `simplified model' means a single WIMP with a single mediator coupling it to the quark sector. This is useful for two reasons; it allows us to better explore which NR operators arise from a broad set of UV complete theories, and also make connection with the growing body of literature which use simplified models for indirect detection and collider searches. 

When it comes to interpreting signals, knowing comprehensively how different interactions with different nuclei arise from different UV complete models will allow us to identify degeneracies between competing models. Further, it can also help optimize target selection for maximum discrimination of the UV model parameter space.

In building these simplified models we remain agnostic about the WIMP's spin, and consider dark matter spins of $0, \half$ and $1$. We do however only consider renormalizable interactions between quarks and WIMPs. To ensure a stable WIMP, we assume that the WIMP is either charged under some internal gauge group or a discrete symmetry group (for example $Z_2$). However, we assume that this gauge charge is not shared by quarks. We will couple the WIMP to the quarks via a heavy mediator in two distinct ways: charged and uncharged mediators, each with all possible spins consistent with angular momentum conservation. The mediator mass is chosen to be the heaviest scale in the problem (and certainly much greater than the momentum exchange which characterizes the scattering process) so that we can integrate it out (see appendix \ref{appNRR} for details). This leads to relativistic effective WIMP-nucleon interactions, whose NR limit can then be examined. In the uncharged mediator case we will consider mediators that are neutral under all SM and WIMP gauge charges, while in the charged case, the mediator must have both WIMP and SM gauge charges. Given the above as a guide, our Lagrangian construction is then constrained only by gauge invariance, Lorentz invariance, renormalizability and hermiticity. In certain cases which follow, the requirement of hermiticity demands coupling constants be complex. Unless explicitly noted, the coupling constants are dimensionless and can be assumed to be real.

\vskip 0.2in

{\bf A. Uncharged-mediator Lagrangians}

{\it 1. Scalar Dark Matter}

We begin with a spin-$0$ scalar WIMP, $S$, which has some internal charge to ensure stability, and $S^\dagger$ is its Hermitian conjugate. To have renormalizable interactions, the neutral mediator can only be a scalar or a vector. We denote the scalar mediator by $\phi$ and the vector mediator by $G^{\mu}$ with field strength tensor $\mathcal{G}_{\mu\nu}$.\\

The most general renormailzable Lagrangian for scalar mediation consistent with the above assumptions is given by
\bea
\mathcal{L}_{S\phi q} &=& \partial_\mu S^\dag\partial^\mu S - m_S^2S^\dag S - \frac{\lambda_S}{2}(S^\dagger S)^2 \nonumber\\
&&+\frac{1}{2}\partial_\mu\phi\partial^\mu\phi - \frac{1}{2}m_\phi^2\phi^2 -\frac{m_\phi\mu_1}{3}\phi^3-\frac{\mu_2}{4}\phi^4 \nonumber\\
&&+ i\bar{q}\Dslash q - m_q \qbar q \nonumber\\
&&-g_1m_SS^\dag S\phi -\frac{g_2}{2}S^\dag S\phi^2-h_1\qbar q\phi-ih_2\bar{q}\gamma^5q\phi, \\ \nonumber
\eea
where we have suppressed all the SM quark interactions. Similarly, the Lagrangian for vector mediation (up to gauge fixing terms) is
\bea
\mathcal{L}_{SGq} \aea \partial_{\mu}S^{\dag}\partial^{\mu}S -m_S^2 S^{\dag}S -\frac{\lambda_S}{2}(S^{\dag}{S})^2  \nonumber\\
&&-\frac{1}{4}\mathcal{G}_{\mu\nu}\mathcal{G}^{\mu\nu} + \frac{1}{2}m_G^2G_{\mu}G^{\mu} -\frac{\lambda_G}{4}(G_{\mu}G^{\mu})^2 \nonumber\\
&&+i\bar{q}\slashed{D}q -m_q\bar{q}q \nonumber\\
&&-\frac{g_3}{2}S^{\dag}SG_{\mu}G^{\mu} -ig_4(S^{\dag}\partial_{\mu}S-\partial_{\mu}S^{\dag}S)G^{\mu}\nonumber\\
&&-h_3(\bar{q}\gamma_{\mu}q	)G^{\mu}-h_4(\bar{q}\gamma_{\mu}\gamma^5q)G^{\mu}.\\ \nonumber
\eea

{\it 2. Spin-$\half$ Dark Matter}

If the WIMP has spin-$\half$ (denoted by $\chi$ below), then, as in the scalar WIMP case, mediation will only occur via scalar or vector mediators. The most general renormalizable interactions for the scalar ($\phi$) and vector mediator ($G_{\mu}$) cases respectively are given below,
\bea
\mathcal{L}_{\chi\phi q} &=& i\bar{\chi}\slashed{D}\chi - m_{\chi}\bar{\chi}\chi \nonumber\\
&&+\frac{1}{2}\partial_\mu\phi\partial^\mu\phi - \frac{1}{2}m_\phi^2\phi^2 -\frac{m_\phi\mu_1}{3}\phi^3-\frac{\mu_2}{4}\phi^4 \nonumber\\
&&+ i\bar{q}\Dslash q - m_q \qbar q \nonumber\\	
&&-\lambda_1\phi\bar{\chi}\chi -i\lambda_2\phi\bar{\chi}\gamma^{5}\chi-h_1\phi\qbar q-ih_2\phi\bar{q}\gamma^5q, \\ \nonumber
\eea
\bea
\mathcal{L}_{\chi Gq} &=& i\bar{\chi}\slashed{D}\chi - m_\chi\bar{\chi}\chi \nonumber\\
&&-\frac{1}{4}\mathcal{G}_{\mu\nu}\mathcal{G}^{\mu\nu}+\frac{1}{2}m_{G}^2G_{\mu}G^{\mu}\nonumber\\
&&+ i\bar{q}\Dslash q - m_q \qbar q \nonumber\\
&&-\lambda_{3}\bar\chi\gamma^\mu\chi G_{\mu}-\lambda_{4}\bar\chi\gamma^\mu\gamma^5\chi G_{\mu}\nonumber\\
&&-h_3\bar{q}\gamma_{\mu}qG^{\mu}-h_4\bar{q}\gamma_{\mu}\gamma^{5}qG^{\mu}.\\ \nonumber
\eea

{\it 3. Spin-$1$ Dark Matter}

If the WIMP is a massive spin-$1$ particle, uncharged mediation to the quark sector can occur via a heavy scalar or a vector particle. For the case of vector mediation, there are many possible interactions because the Lorentz indices on the vectors afford a more diverse set of terms. The general interaction Lagrangian for the scalar mediation case is
\bea
\mathcal{L}_{X\phi q}&=&-\frac{1}{2}{\mathcal{X}}_{\mu\nu}^{\dag}\mathcal{X}^{\mu\nu}+m_{X}^2X_{\mu}^{\dag}X^{\mu}-\frac{\lambda_{X}}{2}(X_{\mu}^{\dag}X^{\mu})^2 \nonumber\\
&&+\frac{1}{2}(\partial_{\mu}\phi)^2-\frac{1}{2}m_{\phi}^2\phi^2-\frac{m_\phi \mu_1}{3}\phi^3-\frac{\mu_2}{4}\phi^4 \nonumber\\
&&+i\bar{q}\slashed{D}q-m_{q}\bar{q}q \nonumber\\
&&-b_1m_X\phi X_{\mu}^{\dag}X^{\mu}-\frac{b_{2}}{2}\phi^2X_{\mu}^{\dag}X^{\mu} -h_1\phi\bar{q}q-ih_2\phi\bar{q}\gamma^{5}q. \\ \nonumber
\eea
For the case of vector mediation, there are many possible interactions because the Lorentz indices on the vectors afford a more diverse set of terms. The Lagrangian is given by
\bea
\mathcal{L}_{XGq}&=& -\frac{1}{2}\mathcal{X}^{\dagger}_{\mu\nu}\mathcal{X}^{\mu\nu}+m_{X}^2X^{\dagger}_{\mu}X^{\mu}-\frac{\lambda_{X}}{2}(X_{\mu}^{\dagger}X^{\mu})^2 \nonumber\\
&&-\frac{1}{4}\mathcal{G}_{\mu\nu}\mathcal{G}^{\mu\nu}+\frac{1}{2}m_{G}^2G_\mu^2-\frac{\lambda_G}{4}(G_\mu G^\mu)^2 \nonumber\\
&&+i\bar{q}\slashed{D}q-m_{q}\bar{q}q\nonumber\\
&&-\frac{b_3}{2}G_\mu^2(X^{\dagger}_\nu X^{\nu}) -\frac{b_{4}}{2}(G^{\mu}G^{\nu})(X^{\dagger}_{\mu}X_{\nu})-\left[ib_{5}X_{\nu}^{\dagger}\partial_{\mu}X^{\nu}G^\mu\right.\nonumber\\
&&\left.+b_{6}X_{\mu}^{\dagger}\partial^\mu X_{\nu}G^{\nu} +b_{7}\epsilon_{\mu\nu\rho\sigma}(X^{\dag\mu}\partial^{\nu}X^{\rho})G^{\sigma} +h.c.\right]\nonumber\\
&&-h_3G_\mu\bar{q}\gamma^\mu q - h_4 G_\mu\bar{q}\gamma^\mu\gamma^{5}q \\ \nonumber
\eea
where, for the Lagrangian to be Hermitian, $b_6$ and $b_7$ are complex (this implies a new source of CP violation, which will not be considered further here).

\subsection{Charged-mediator Lagrangians}
Here we consider the simplest case of mediators that are charged under both the DM internal symmetry group and SM gauge groups. This is motivated by the absence of spin-$\half$ mediators ($s$-channel processes) in the previous section. Such a mediator, if neutral, is forbidden by simultaneous requirements of gauge invariance and renormalizability. Dark Matter models with mediators endowed with charges from both DM and SM side have been considered in the literature before \cite{Hisano:2010yh,Feng:2008ya}. The case of a spin-$\half$ mediator carrying $SU(3)_c$ is also motivated by studies of heavy quark models. This allows unique interactions as we show below.  In particular they necessitate a direct interaction between quarks and WIMPs at the level of the Lagrangian.

{\it 1. Scalar Dark Matter}

Scalar WIMPs with a charged scalar or vector mediator do not lead to any Lorentz invariant interactions. This is easy to see since both the scalars (or scalar and vector) and the quark are required in the (gauge invariant) interaction, but there is no way to contract the spinor indices consistently if the mediating particle is a scalar or vector. Therefore, the only possibility is that of a spin-$1/2$ mediator, $Q$, which acts like a heavy quark. The general renormalizable action is given by
\bea
\mathcal{L}_{SQq} \aea \partial_{\mu}S^{\dag}\partial^{\mu}S -m_S^2 S^{\dag}S -\lambda_S(S^{\dag}{S})^2 \nonumber\\
&&+i\bar{Q}\slashed{D}Q-m_Q\bar{Q}Q\nonumber\\
&&+i\bar{q}\slashed{D}q -m_q\bar{q}q\nonumber\\
&&-(y_1S\bar{Q}q+y_2S\bar{Q}\gamma^5q +h.c.),\\\nonumber
\eea
where $y_1$ and $y_2$ are again complex.

{\it 2. Spin-$\half$ Dark Matter}

For a spin-$1/2$ WIMP, both a charged scalar and charged vector mediator exchange can lead to novel interactions. The charged scalar is denoted by $\Phi$ and the charged vector by $V_\mu$ 
\bea
\mathcal{L}_{\chi\Phi q} \aea i\bar{\chi}\slashed{D}\chi-m_{\chi}\bar{\chi}\chi\nonumber\\
&&+(\partial_{\mu}\Phi^{\dagger})(\partial^{\mu}\Phi)-m_{\Phi}^2\Phi^{\dagger}\Phi-\frac{\lambda_\Phi}{2}(\Phi^{\dagger}\Phi)^2\nonumber\\
&&+i\bar{q}\slashed{D}q-m_{q}\bar{q}q \nonumber\\
&&-(l_{1}\Phi^{\dagger}\bar{\chi}q+l_{2}\Phi^{\dagger}\bar{\chi}\gamma^{5}q+h.c.),\\ \nonumber
\eea \vspace{-2cm}
\bea
\mathcal{L}_{\chi Vq} \aea i\bar{\chi}\slashed{D}\chi-m_{\chi}\bar{\chi}\chi\nonumber\\
&&-\frac{1}{2}\mathcal{V}_{\mu\nu}^{\dagger}\mathcal{V}^{\mu\nu}+m_{V}^2V^{\dagger}_{\mu}V^{\mu}\nonumber\\
&&+i\bar{q}\slashed{D}q-m_{q}\bar{q}q\nonumber\\
&&-(d_{1}\bar{\chi}\gamma^{\mu}qV^{\dagger}_{\mu}+d_{2}\bar{\chi}\gamma^{\mu}\gamma^{5}qV^{\dagger}_{\mu}+h.c.),\\ \nonumber
\eea
where $l_1, l_2, d_1$ and $d_2$ are complex.

{\it 3. Vector DM}

Here again we only have the case of a spin-$\half$ mediated interaction between vector DM and quarks (again scalar and vector charged mediators aren't possible because they don't lead to Lorentz invariant and renormalizable interactions). The general Lagrangian is given by 
\bea
\mathcal{L}_{XQq} \aea-\frac{1}{2}\mathcal{X}^{\dagger}_{\mu\nu}\mathcal{X}^{\mu\nu}+m_{X}^2X^{\dagger}_{\mu}X^{\mu}-\frac{\lambda_{X}}{2}(X_{\mu}^{\dagger}X^{\mu})^2 \nonumber\\
&&+i\bar{Q}\slashed{D}Q-m_Q\bar{Q}Q\nonumber\\
&&+i\bar{q}\slashed{D}q-m_{q}\bar{q}q\nonumber\\
&&-(y_3X_{\mu}\bar{Q}\gamma^{\mu}q + y_4X_{\mu}\bar{Q}\gamma^{\mu}\gamma^5q+h.c.),\\\nonumber
\eea
where $y_3$ and $y_4$ are complex.

\section{Non-relativistic reduction of simplified models}
\label{secNRR}
After integrating out the heavy mediator we replace quark operators with nucleon operators (see appendix \ref{appQ2N}), take the non-relativistic limit (see appendix \ref{appNRR}), and match onto the operators given in table~\ref{tabHaxtonOp}. The results of this calculation are presented in terms of the $c_i$ coefficients from \cite{Anand:2013yka}, described in section \ref{secEFT}, facilitating a straightforward computation of amplitudes and rates. The $c_i$'s are given for each of the WIMP spins in tables \ref{tabCiScalar}, \ref{tabCiSpinor} and \ref{tabCiVector}. With this general framework in place we can now easily find the leading order NR operators for each distinct WIMP-nucleus interaction. One can imagine a series of minimal scenarios in which a combination of two Lagrangian couplings that give rise to a direct detection signal is non-zero with all others set to zero, and then proceeding in this manner for the entire set. Each of these scenarios is listed with its leading operators in table~\ref{tabLeading} and with all operators generated in table~\ref{tabFull}. Note that in the case of a complex coupling constant we consider purely real and purely imaginary values as separate cases since they produce a distinct set of operators.
\begin{table}[hbt]
\caption{Non-zero $c_i$ coefficients for a spin$-0$ WIMP}
\begin{tabular}{|c|c|c|}
\hline
 & Uncharged Mediator & Charged Mediator \\
\hline
 $c_{1}$ & $\frac{h_1^Ng_{1}}{m_{\phi}^2}$ & $\frac{y_1^\dag y_1-y_2^\dag y_2}{m_Q m_S}f_{T}^N$\\ 
$c_{10}$ & $\frac{-ih_2^Ng_{1}}{m_{\phi}^2} + \frac{2ig_4h_4^N}{m_G^2}\frac{m_N}{m_S} $ & $i\frac{y_2^\dag y_1 - y_1^\dag y_2}{m_Q m_S}\tilde\Delta^N$\\
\hline
\end{tabular}
\label{tabCiScalar}
\end{table}

\begin{table}[htb]
\caption{$c_i$ coefficients for a spin-$\half$ WIMP}
\begin{tabular}{|c|c|c|}
\hline
 & Uncharged Mediator & Charged Mediator \\
\hline
 $c_{1}$ & $\frac{h_1^N \lambda_1}{m_\phi^2}  - \frac{h_3^N\lambda_3}{m_G^2}$ & $\left(\frac{l_2^\dag l_2-l_1^\dag l_1}{4 m_\Phi^2}+\frac{d_2^\dag d_2-d_1^\dag d_1}{4m_V^2}\right)f_{T}^N+\left(-\frac{l_2^\dag l_2+l_1^\dag l_1}{4m_\Phi^2} +\frac{d_2^\dag d_2+d_1^\dag d_1}{8m_V^2}\right)\mathcal{N}^N$\\
 $c_{4}$ & $\frac{4h_4^N \lambda_4}{m_G^2} $ & $\frac{l_2^\dag l_2-l_1^\dag l_1}{m_\Phi^2}\delta^N-\left(\frac{l_1^\dag l_1+l_2^\dag l_2}{m_\Phi^2}+\frac{d_2^\dag d_2-d_1^\dag d_1}{2m_V^2}\right)\Delta^N$\\
 $c_{6}$ & $\frac{h_2^N \lambda_2m_N}{m_\phi^2m_\chi}$ & $(\frac{l_1^\dag l_1-l_2^\dag l_2}{4m_\Phi^2}+\frac{d_2^\dag d_2-d_1^\dag d_1}{4m_V^2})\frac{m_N}{m_\chi}\tilde{\Delta}^N$\\
 $c_{7}$ & $\frac{2h_4^N\lambda_3}{m_G^2} $& $(\frac{l_1^\dag l_2-l_2^\dag l_1}{2m_\Phi^2}+\frac{d_1^\dag d_2+d_2^\dag d_1}{4m_V^2})\Delta^N$\\
 $c_{8}$ & $-\frac{2h_3^N\lambda_4}{m_G^2} $& $(\frac{l_1^\dag l_2-l_2^\dag l_1}{2m_\Phi^2}-\frac{d_1^\dag d_2+d_2^\dag d_1}{4m_V^2})\mathcal{N}^N$\\
 $c_{9}$ & $-\frac{2h_4^N\lambda_3 m_N}{m_\chi m_G^2} -\frac{2h_3^N\lambda_4}{m_G^2} $ & $(\frac{l_1^\dag l_2-l_2^\dag l_1}{2m_\Phi^2}-\frac{d_1^\dag d_2+d_2^\dag d_1}{4m_V^2})\mathcal{N}^N-(\frac{l_1^\dag l_2-l_2^\dag l_1}{2m_\Phi^2}-\frac{d_1^\dag d_2+d_2^\dag d_1}{4m_V^2})\frac{m_N}{m_\chi}\Delta^N$\\
$c_{10}$ & $\frac{h_2^N \lambda_1}{m_\phi^2}  $& $i(\frac{l_1^\dag l_2-l_2^\dag l_1}{4 m_\Phi^2} + \frac{d_2^\dag d_1-d_1^\dag d_2}{4 m_V^2})\tilde\Delta^N-i\frac{l_1^\dag l_2-l_2^\dag l_1}{m_\Phi^2}\delta^N$\\
$c_{11}$ & $-\frac{h_1^N \lambda_2m_N}{m_\phi^2 m_\chi} $ & $i(\frac{l_2^\dag l_1-l_1^\dag l_2}{4 m_\Phi^2} + \frac{d_2^\dag d_1-d_1^\dag d_2}{4 m_V^2})\frac{m_N}{m_\chi}f_{T}^N+i\frac{l_1^\dag l_2-l_2^\dag l_1}{m_\Phi^2}\frac{m_N}{m_\chi}\delta^N$\\
$c_{12}$ & 0 & $\frac{l_2^\dag l_1-l_1^\dag l_2}{m_{\Phi}^2}\delta^N$\\
\hline
\end{tabular}
\label{tabCiSpinor}
\end{table}

\begin{table}[htb]
\caption{$c_i$ coefficients for a spin-$1$ WIMP}
\begin{tabular}{|c|c|c|}
\hline
 & Uncharged Mediator & Charged Mediator \\
\hline
$c_{1}$ & $\frac{b_1h_1^N}{m_\phi^2}$ & $\frac{y_3^\dag y_3-y_4^\dag y_4}{m_Q m_X}f^N_T$\\
$c_{4}$ & $\frac{4\mathrm{Im}(b_7)h_4^N}{m_G^2} + i\frac{q^2}{m_X^2}\frac{\mathrm{Re}(b_7)h_4^N}{m_G^2}-\frac{q^2}{m_Xm_N}\frac{\mathrm{Re}(b_6)h_3^N}{m_G^2}$ & $2\frac{y_3^\dag y_3-y_4^\dag y_4}{m_Q m_X}\delta^N$\\
$c_{5}$ & $\frac{\mathrm{Re}(b_6)h_3^N}{m_G^2}\frac{m_N}{m_X}$  & 0\\
$c_{6}$ & $\frac{\mathrm{Re}(b_6)h_3^N}{m_G^2}\frac{m_N}{m_X}-i\frac{\mathrm{Re}(b_7)h_4^N}{m_G^2}\frac{m_N^2}{m_X^2}$  & 0\\
$c_{8}$ & $ \frac{2\mathrm{Im}(b_7)h_3^N}{m_G^2}$ & 0\\
$c_{9}$ & $-\frac{2\mathrm{Re}(b_6)h_4^N}{m_G^2}\frac{m_N}{m_X}+\frac{2\mathrm{Im}(b_7)h_3^N}{m_G^2}$ & 0\\
$c_{10}$& $ \frac{b_1h_2^N}{m_\phi^2}-\frac{3b_5h_4^N}{m_G^2}\frac{m_N}{m_X}$&  $i\frac{y_4^\dag y_3-y_3^\dag y_4}{m_Q m_X}\tilde\Delta^N$\\
$c_{11}$ & $ \frac{\mathrm{Re}(b_7)h_3^N}{m_G^2}\frac{m_N}{m_X}$&  $i\frac{y_4^\dag y_3-y_3^\dag y_4}{m_Q m_X}\delta^N$\\
$c_{12}$ &0& $2i\frac{y_3^\dag y_4-y_4^\dag y_3}{m_Q m_X}\delta^N$\\
$c_{14}$ & $-\frac{2\mathrm{Re}(b_7)h^N_4}{m_G^2}\frac{m_N}{m_X}$& 0\\
$c_{17}$ & $-\frac{4\mathrm{Im}(b_6)h^N_3}{m_G^2}\frac{m_N}{m_X}$& 0\\
$c_{18}$ & $ \frac{4\mathrm{Im}(b_6)h^N_4}{m_G^2}\frac{m_N}{m_X}$& $-2i\frac{y_4^\dag y_3-y_3^\dag y_4}{m_Q m_X}\delta^N$\\
\hline
\end{tabular}
\label{tabCiVector}
\end{table}

As described earlier, we find that it is important to consider operators beyond those incorporated into the standard spin-independent and spin-dependent formalism, i.e.~simple models exist in which one would infer an incorrect rate in current experiments by not including these effects. Also importantly, not all of the NR operators are actually generated at leading order; for example, the operators $\mathcal{O}_2$, $\mathcal{O}_3$, $\mathcal{O}_{13}$ and $\mathcal{O}_{15}$ are missing at leading order.  Note that we only consider renormalizable Lagrangians, higher order non-renormalizable operators, which are presumably further suppressed. We have also not considered the case of kinetic mixing, which could be used to generate anapole interactions~\cite{Gresham:2014vja}, because the effective interaction doesn't arise from one mediator exchange.  \\

While spin independent interactions are a generic feature of direct couplings to quarks in our charged mediator cases, it is sometimes possbile to suppress them. In the scalar (and vector) WIMP with charged mediator cases, it is possible to suppress the spin independent interaction by ensuring that $|y_1|=|y_2|$($|y_3|=|y_4|$) while keeping their relative phases non-zero (or $\pi$). While these non-minimal scenarios require some fine tuning we include it for completeness and label them $y_1,y_2$ and $y_3,y_4$.\\

Aside from scalar WIMPs, each particular spin produces some non-relativistic operators that are unique to that spin. Also, importantly, the operators $\mathcal{O}_1$ and $\mathcal{O}_{10}$ are generic to all spins.  In five cases relativistic operators generate unique non-relativistic operators at leading order.  Therefore distinguishing WIMP scenarios in these cases reduces to experimentally discerning between these operators (see also \cite{catena:2014}).  Given the likely low statistics of any detection in upcoming direct detection experiments, sub-leading operators are not likely to contribute enough to provide any further discriminating power. \\

\begin{table}[htb]
\renewcommand{\arraystretch}{0.9}
\caption{Leading order operators which can arise from the relativistic Lagrangians considered in this work, the column `$\mathcal{L}$ terms' gives the non-zero couplings for that scenario. Each row represents a possible leading order direct detection signal. A `$\dag$' indicates that the mediator is charged. The 'Eqv.~$M_m$' column gives the mediator mass required for each scenario to produce $\sim$10 events $t^{-1}yr^{-1}keV^{-1}$ in xenon, with couplings set to 0.1. }
\begin{tabular}{|c|c|c|c|r|}
\hline
WIMP spin & Mediator spin & $\mathcal{L}$ terms& leading NR operator & Eqv.~$M_m$ \\
\hline
0        & 0 			& 	$h_1,g_1$      & $\mathcal{O}_1$	 & 13 TeV\\
0  	   & 0 			& 	$h_2,g_1$      & $\mathcal{O}_{10}$ & 14 GeV\\
0		   & 1 			& 	$h_4,g_4$      & $\mathcal{O}_{10}$ & 8 GeV\\
0 		   & $\half^\dag$ &	$y_1$          & $\mathcal{O}_{1}$  & 3.2 PeV\\
0 		   & $\half^\dag$ &	$y_2$          & $\mathcal{O}_{1}$  & 3.2 PeV\\
0 		   & $\half^\dag$ &	$y_1,y_2$      & $\mathcal{O}_{10}$ & 41 GeV\\
\hline
$\half$  & 0			& $h_1,\lambda_1$ & $\mathcal{O}_1$    & 12.7 TeV\\
$\half$  & 0			& $h_2,\lambda_1$ & $\mathcal{O}_{10}$ & 293 GeV\\
$\half$  & 0			& $h_1,\lambda_2$ & $\mathcal{O}_{11}$ & 14 GeV\\
$\half$  & 0			& $h_2,\lambda_2$ & $\mathcal{O}_6$    & 1.9 GeV\\
$\half$  & 1	   	& $h_3,\lambda_3$ & $\mathcal{O}_1$    & 6.3 TeV\\
$\half$  & 1		   & $h_4,\lambda_3$ & $\mathcal{O}_9$    & 6.4 GeV\\
$\half$  & 1		   & $h_3,\lambda_4$ & $\mathcal{O}_8$    & 180 GeV\\
$\half$  & 1		   & $h_4,\lambda_4$ & $\mathcal{O}_4$    & 135 GeV\\
$\half$	& $0^\dag$			& $l_1$		 		 & $\mathcal{O}_1$   & 7.1 TeV\\
$\half$	& $0^\dag$			& $l_2$		 		 & $\mathcal{O}_1$   & 5.5 TeV\\
$\half$	& $1^\dag$			& $d_1$		 		 & $\mathcal{O}_1$   & 5.9 TeV\\
$\half$	& $1^\dag$			& $d_2$		 		 & $\mathcal{O}_1$   & 6.7 TeV\\
\hline
1 			&  0			& $h_1,b_1$ 		& $\mathcal{O}_1$     & 13 TeV\\
1 			&  0		   & $h_2,b_1$ 		& $\mathcal{O}_{10}$  & 10 GeV\\
1 			&  1			& $h_4,b_5$ 		& $\mathcal{O}_{10}$  & 5.1 GeV\\
1 			&  1			& $h_3,b_6^\mathrm{Re}(b_6^\mathrm{Im})$ 		& $\mathcal{O}_{5}(\mathcal{O}_{17})$  & 5.5 GeV(23 GeV)\\
1 			&  1			& $h_4,b_6^\mathrm{Re}(b_6^\mathrm{Im})$ 		& $\mathcal{O}_{9}(\mathcal{O}_{18})$  & 3 GeV(4.6 GeV)\\
1 			&  1			& $h_3,b_7^\mathrm{Re}(b_7^\mathrm{Im})$ 		& $\mathcal{O}_{11}(\mathcal{O}_{8})$  & 186 GeV(228 GeV)\\
1 			&  1			& $h_4,b_7^\mathrm{Re}(b_7^\mathrm{Im})$ 		& $\mathcal{O}_{4}(\mathcal{O}_{4})$     & 78 MeV (172 GeV)\\
1		   &$\half^\dag$ & $y_3$		 		& $\mathcal{O}_1$     & 3.2 PeV\\
1  		&$\half^\dag$ & $y_4$		 		& $\mathcal{O}_1$     & 3.2 PeV\\
1  		&$\half^\dag$ & $y_3,y_4$			& $\mathcal{O}_{11}$  & 120 TeV\\
\hline
\end{tabular}
\label{tabLeading}
\end{table}

\section{Observables}

\begin{figure}
\includegraphics[height=4.6cm]{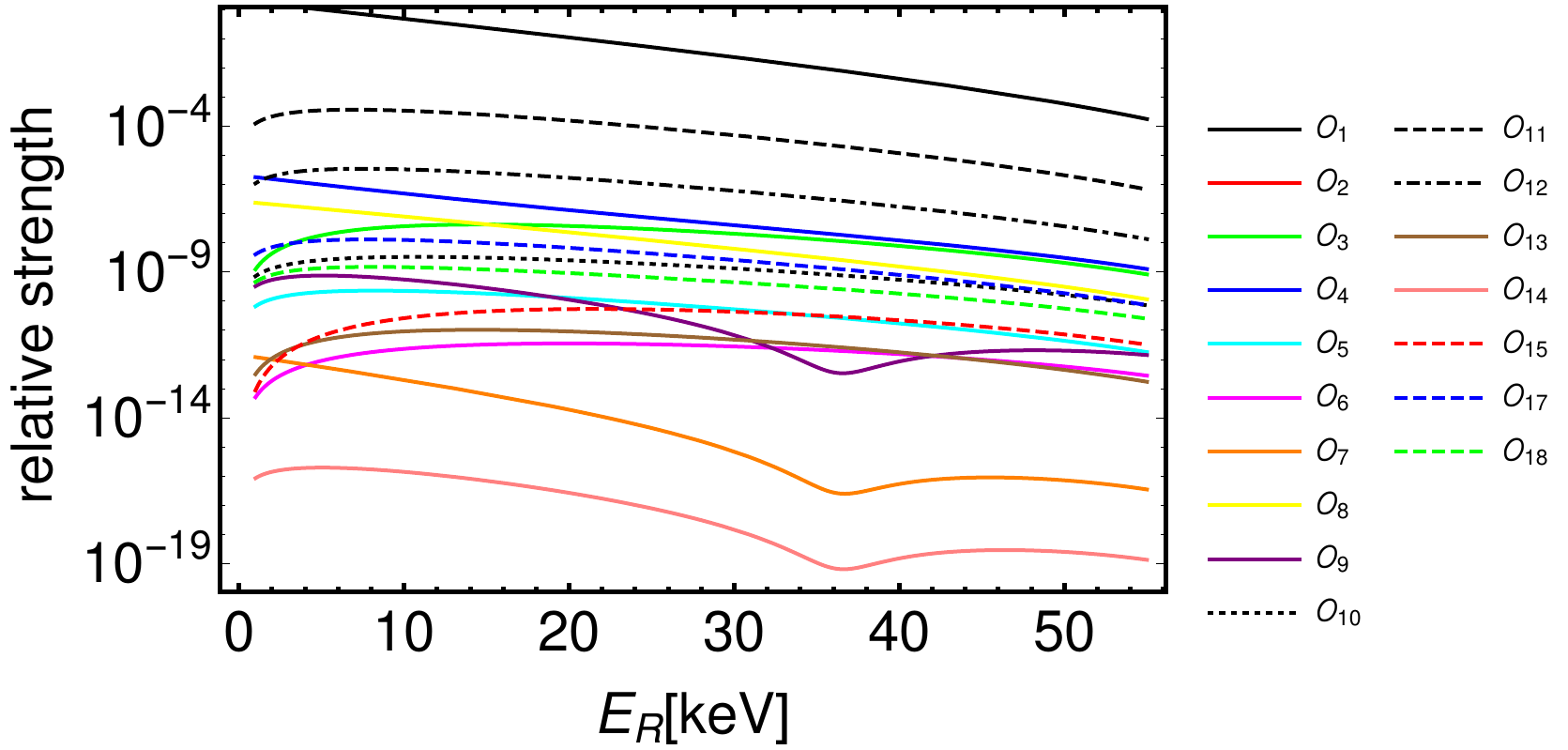}
\caption{The relative strength of event rates for a 50GeV spin-$\half$ WIMP in xenon for each of the non-relativistic operators in table \ref{tabHaxtonOp}, where the coefficients of each operator are set to be equal}
\label{figOperators}
\end{figure}

The principle observable in direct detection experiments is the differential event rate. Since the incoming WIMPs originate in the galactic halo, one must average over the WIMP velocity distribution, $f(v)$, which we assume for the purposes of this paper to be Maxwell-Boltzmann,
\bea
\frac{dR}{dE_R} = N_T \frac{\rho_\chi M}{2\pi m_\chi}\int_{v_{min}}\frac{f(v)}{v} P_{tot} dv
\eea
where we use the value $\rho_\chi=0.3$GeV/cm$^3$ for the local dark matter density, $N_T$ is the number of nuclei in the target and $P_{tot}$ can be calculated from the amplitude $\mathcal{M}$ in Eq.~\ref{eqnAmp}
\bea
P_{tot} = \frac{1}{2j_\chi+1}\frac{1}{2j_N+1}\sum_{spins}|\mathcal{M}|^2.
\eea
Thoughout this work we use the mathematica package supplied in~\cite{Anand:2013yka} to calculate rates. To determine the leading order operator which arises from a given relativistic scenario we first plot the rate for each of the NR operators in xenon-131. To simply compare the operators we set the $c_i$ coefficients to be the same and normalized the overall rate to that of $\mathcal{O}_1$, see Fig.~\ref{figOperators}. Since operators are either zero, first or second order in momentum transfer $q$ or velocity $\vec{v}^\perp$, the relative strengths of the operators span 16 orders of magnitude. This is an important point to keep in mind when finding the leading operator, as sometimes a term which appears to be higher order in $q$ can dominate the non-relativistic reduction. For example in the $b_7^{\mathrm{Re}}h_4$ scenario, one finds that $q^2\mathcal{O}_4$ dominates over the $\mathcal{O}_6$ and $\mathcal{O}_{14}$ which contain powers of $q$ within the operators. \\

Since the Lagrangians we have considered are not tied to specific complete and consistent particle physics models,  the mediator masses are not fixed in advance and thus specific event rates are not predicted in advance.  Clearly one requires a rate that is low enough to evade the current experimental constraints.  For example, a 50 GeV WIMP producing 10 events per tonne per year is sufficiently low to evade the bounds from LUX~\cite{Akerib:2013tjd}. For demonstration purposes we set the couplings to 0.1 (or $0.1i$ for imaginary) in the various Lagrangians and find a mediator mass that will produce 10 events/t/y in the signal region for xenon ($5-45$keV). The calculated masses are given in table~\ref{tabLeading}.  It is perhaps telling that the mediator masses span 6 orders of magnitude, from just a few GeV up to a PeV.  While it is unlikely that a full model of thermal relic dark matter could be built around all of these Lagrangians, it is nevertheless a useful metric to estimate the relative strength of the different nuclear responses to each of the operators. \\

\begin{figure}[ht]
\mbox{
\includegraphics[width=8cm]{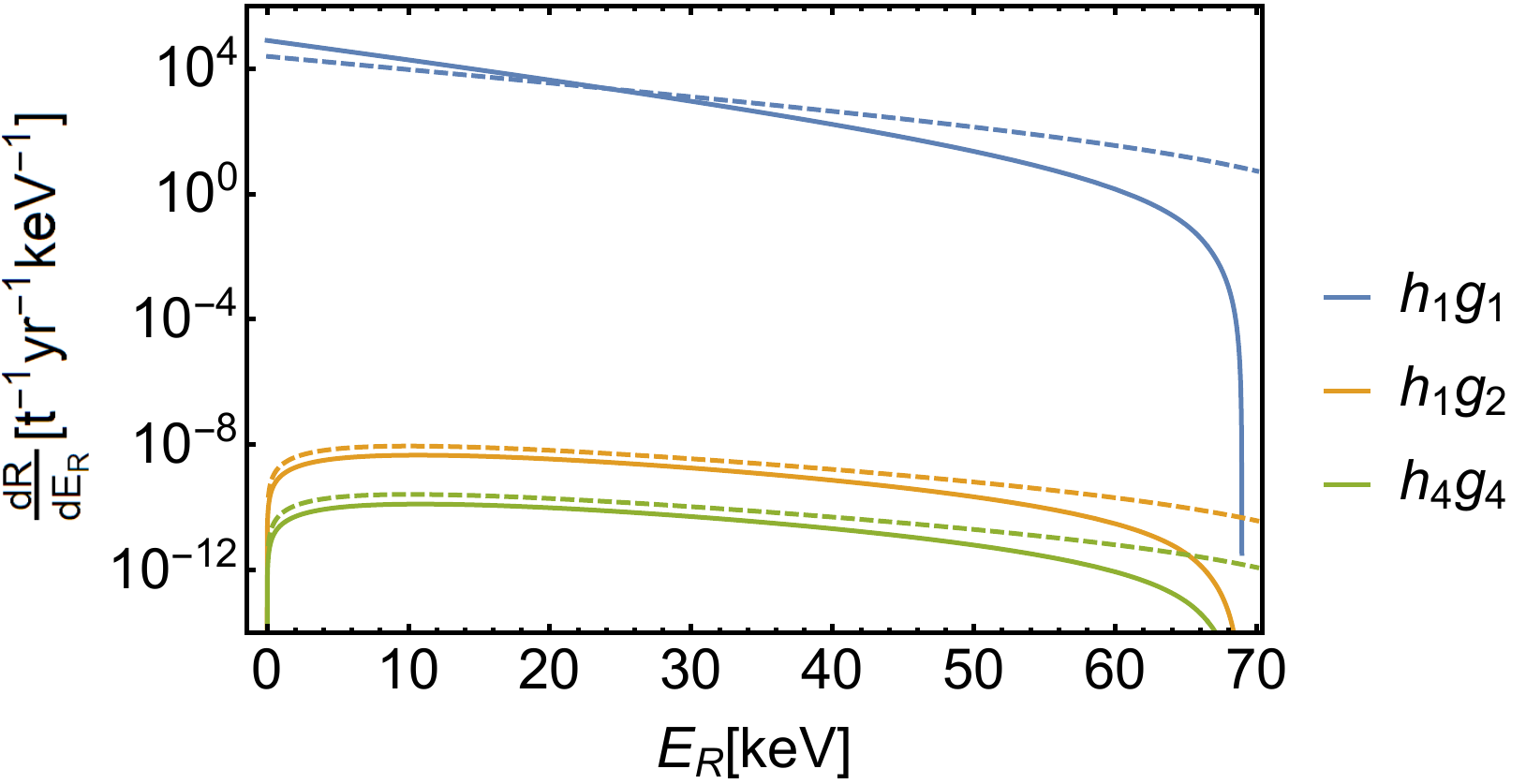}
\includegraphics[width=8cm]{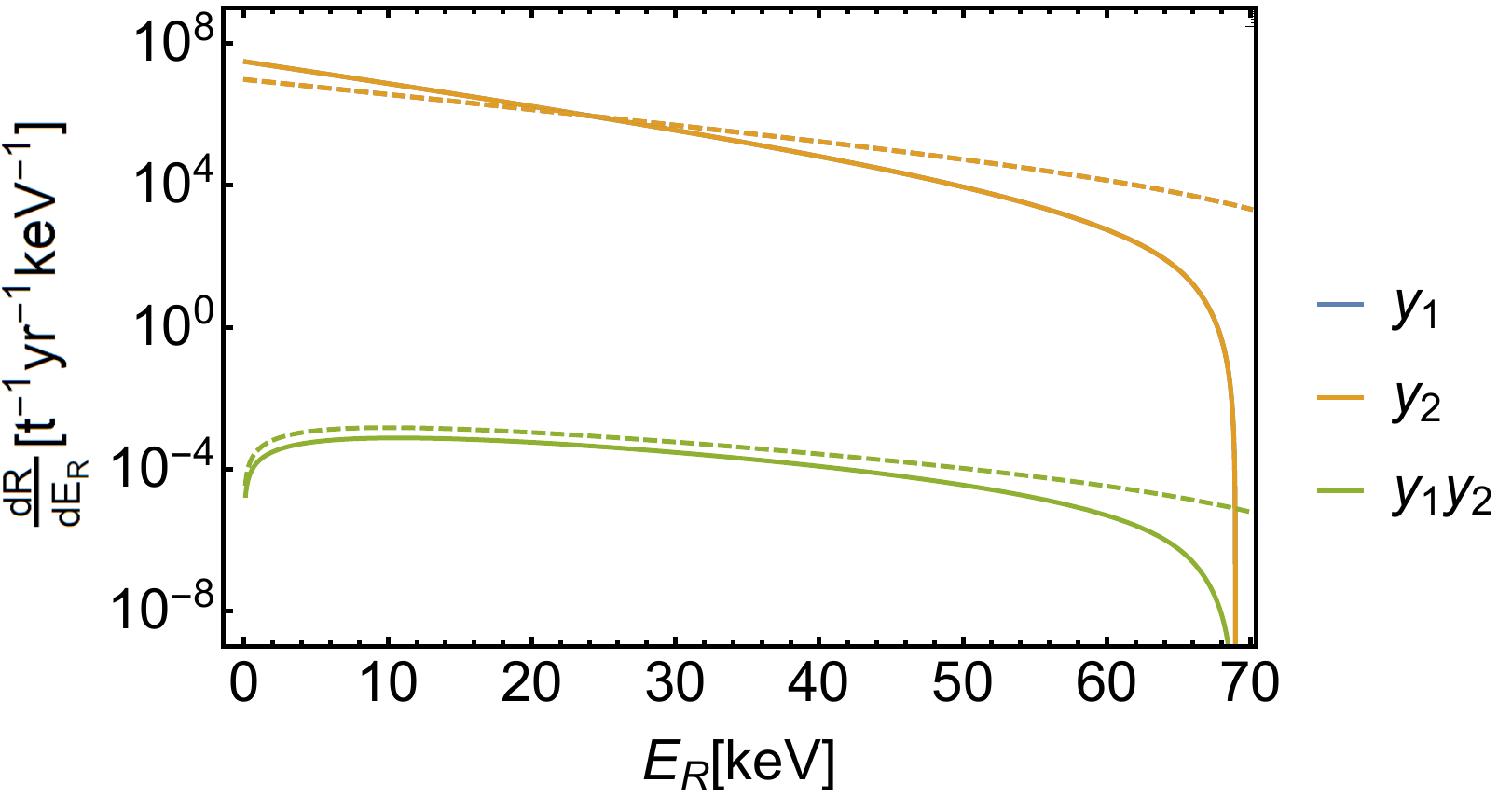}
}
\caption{Rates for a 50GeV spin-0 WIMP in xenon (solid) and germanium (dashed) with uncharged (left) and charged mediators (right), assuming mediator mass of 1TeV and $\mathcal{O}(1)$ coupling constants.}
\label{figScalar}
\end{figure}

\begin{figure}[ht]
\mbox{
\includegraphics[height=4.2cm]{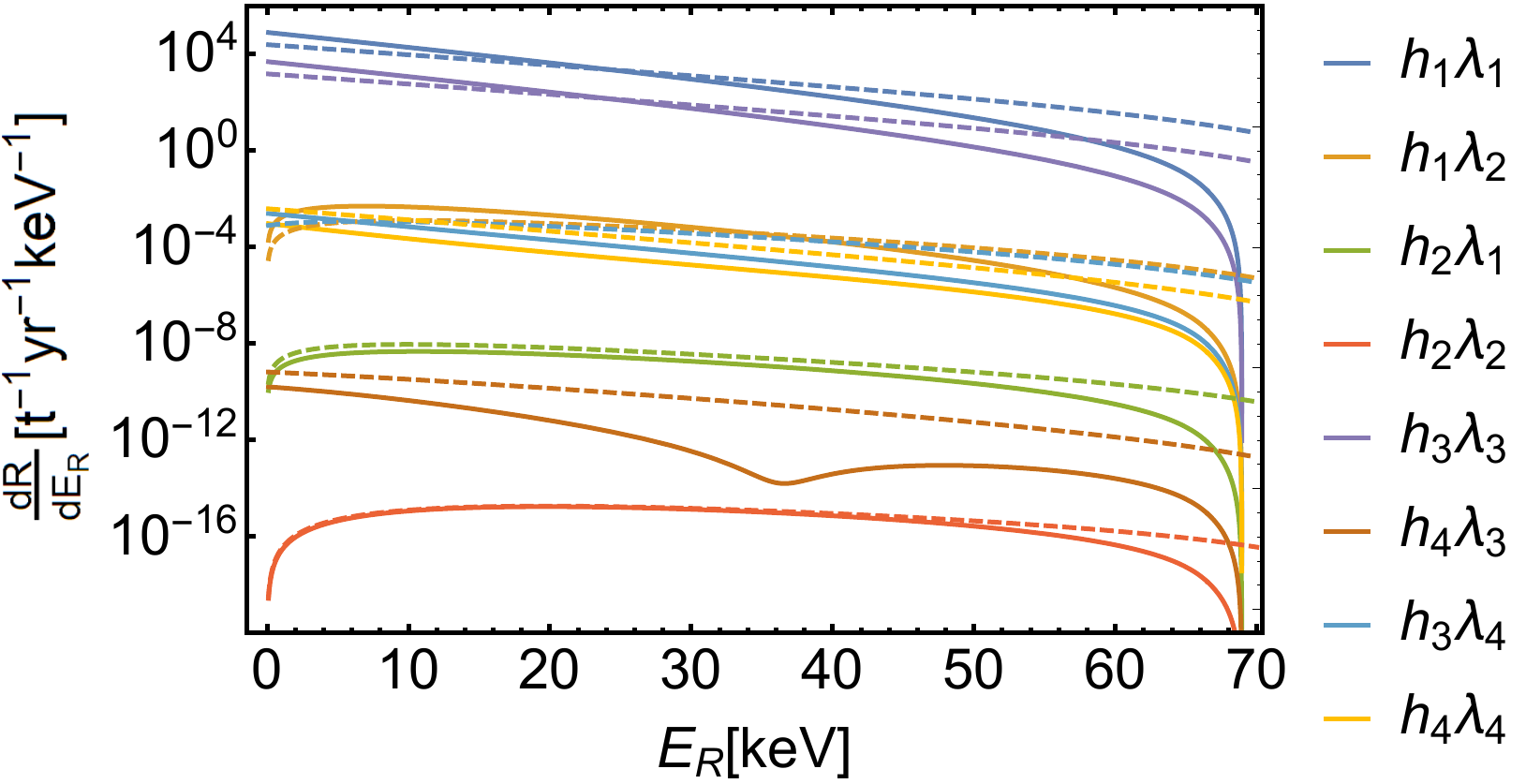}
\includegraphics[height=4.2cm]{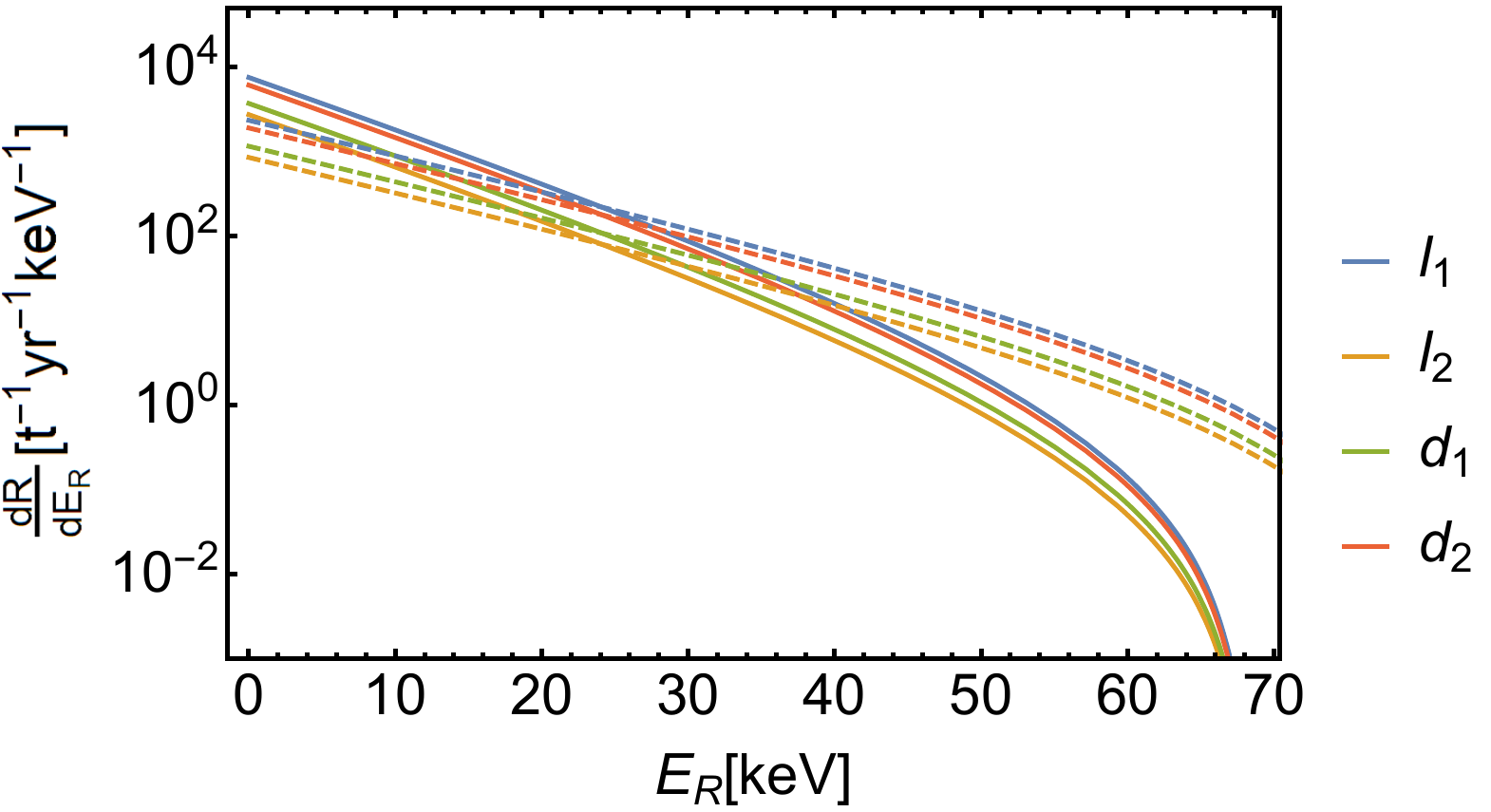}
}
\caption{Rates for a 50GeV spin-$\half$ WIMP in xenon (solid) and germanium (dashed) with uncharged (left) and charged mediators (right), assuming mediator mass of 1TeV and $\mathcal{O}(1)$ coupling constants.}
\label{figSpinor}
\end{figure}

\begin{figure}[ht]
\mbox{
\includegraphics[height=4.2cm]{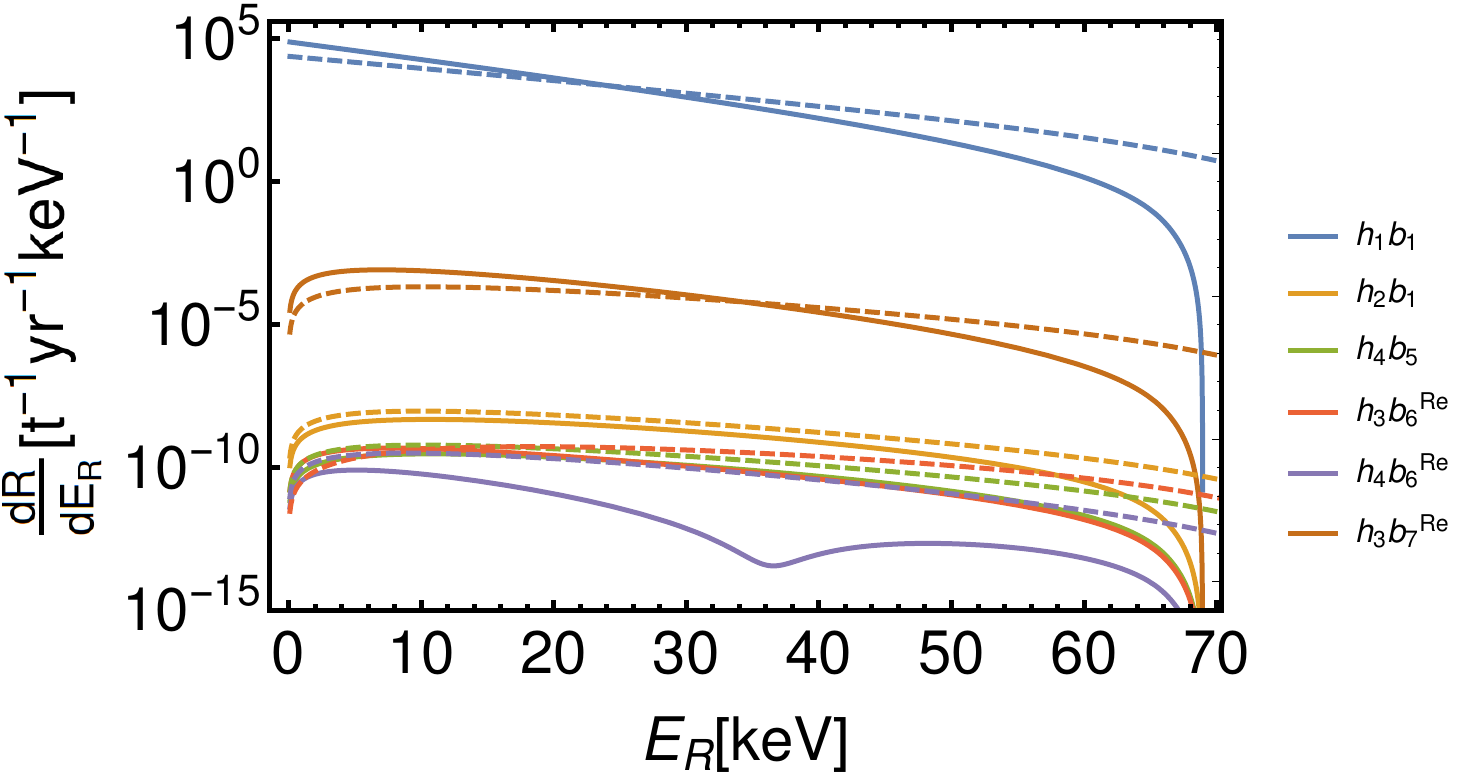}
\includegraphics[height=4.2cm]{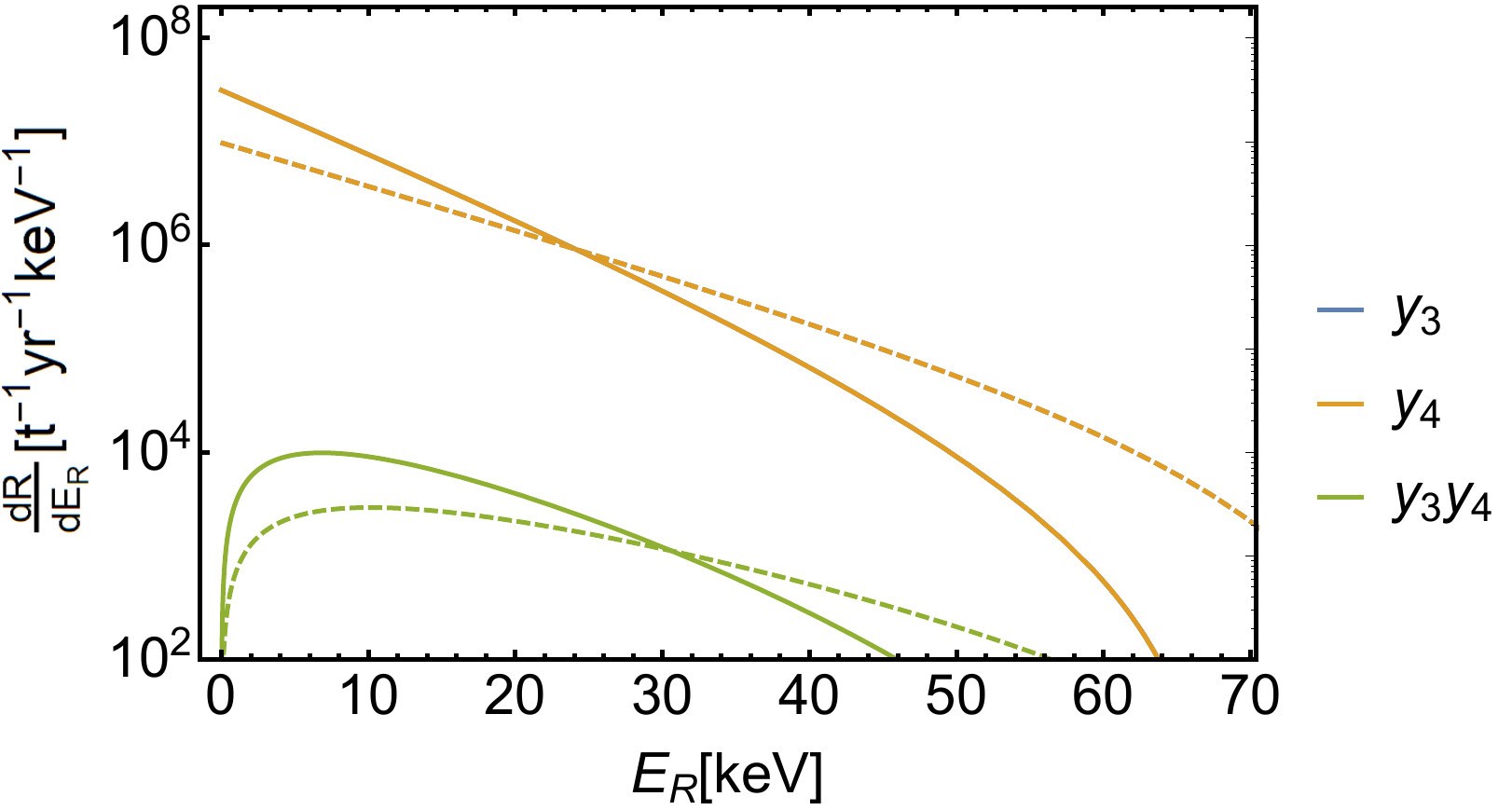}
}
\caption{Rates for a 50GeV spin-1 WIMP in xenon (solid) and germanium (dashed) with uncharged (left) and charged mediators (right), assuming mediator mass of 1TeV and $\mathcal{O}(1)$ coupling constants.}
\label{figVector}
\end{figure}

\begin{figure}[ht]
\mbox{
\includegraphics[width=8cm]{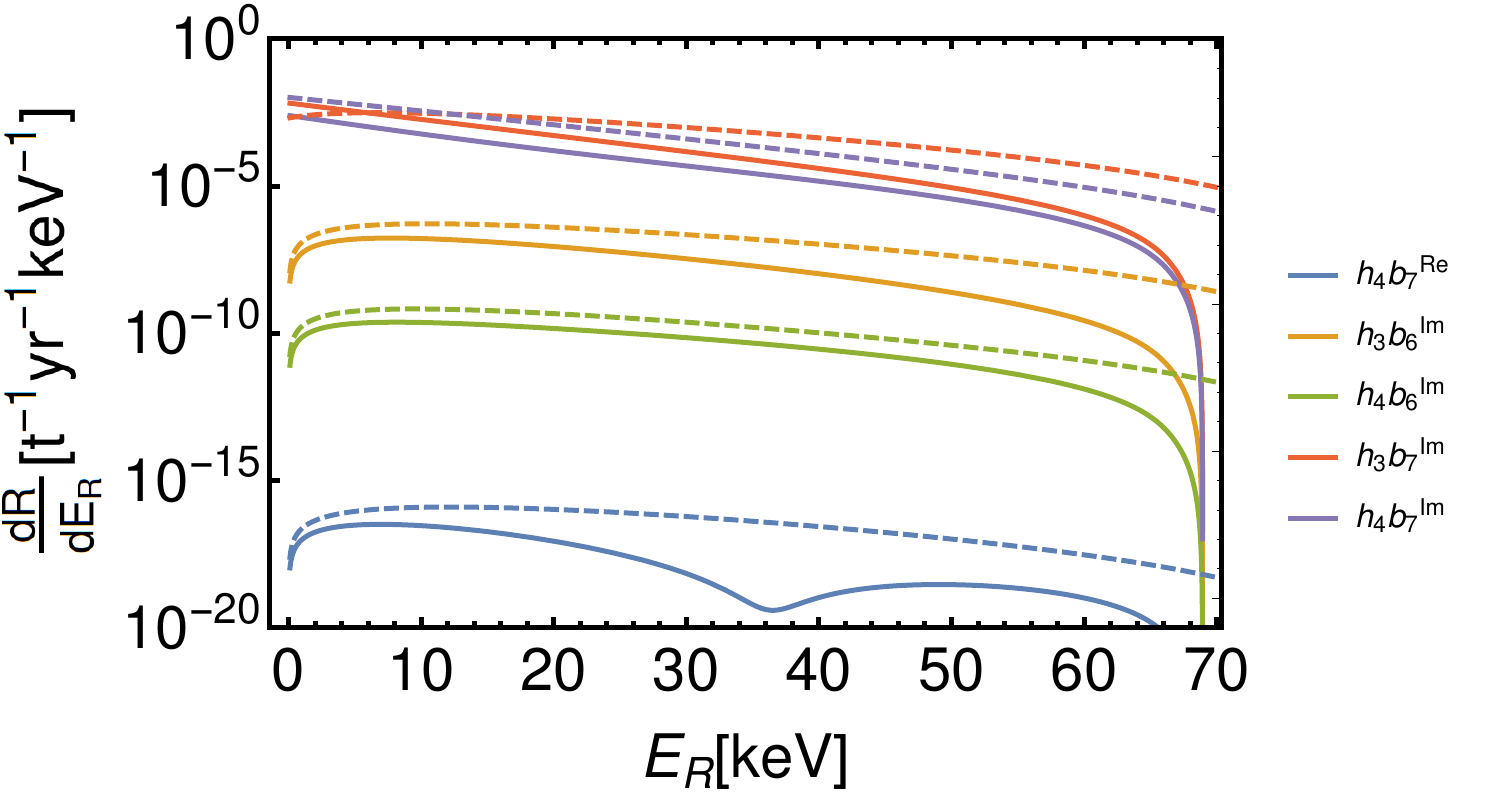}
\includegraphics[width=8cm]{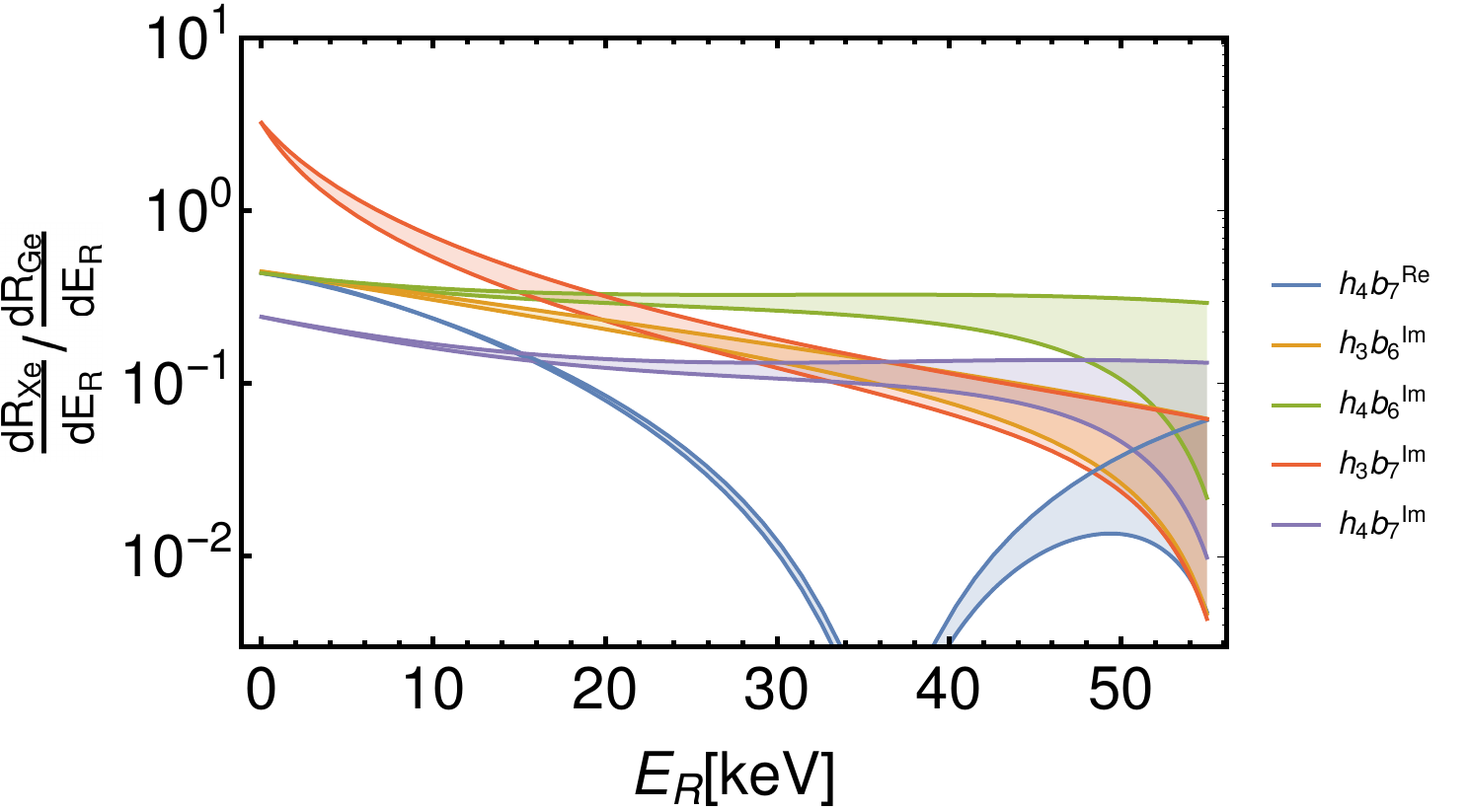}
}
\caption{Rates (left) for a 50GeV spin-1 WIMP in xenon (solid) and germanium (dashed) with uncharged mediators and imaginary couplings, assuming mediator mass of 1TeV and $\mathcal{O}(1)$ coupling constants. Also shown is the ratio of rates in xenon and germanium (right).}
\label{figVectorI}
\end{figure}

In Figs.~\ref{figScalar},~\ref{figSpinor},~\ref{figVector} and ~\ref{figVectorI} we have plotted rates for two common targets. For simplicity and again for demonstration purposes, we only plot the rates for a single isotope of both germanium and xenon. The choice of isotopes, $^{73}$Ge and $^{131}$Xe, was made to ensure sensitivity to spin-dependent responses. As can be seen in the figures, many operators produce rates with similar recoil energy dependence in the same target, but different nuclei can have very different responses to the various operators~\cite{Fitzpatrick:2012ix}. Thus a complementary choice of nuclear targets can provide important discriminating information. 

\begin{figure}[ht]
\mbox{
\includegraphics[width=8cm]{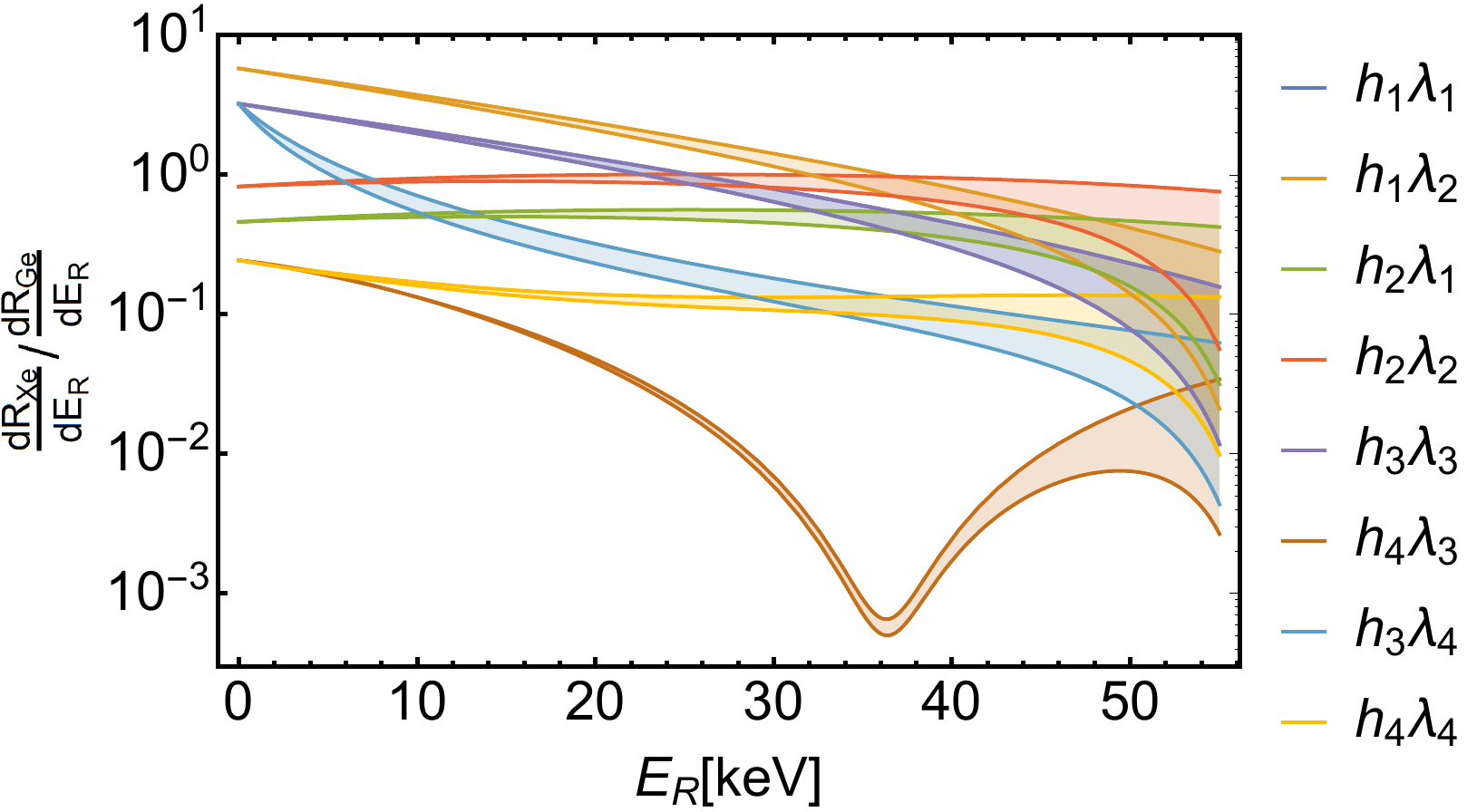}
\includegraphics[width=8cm]{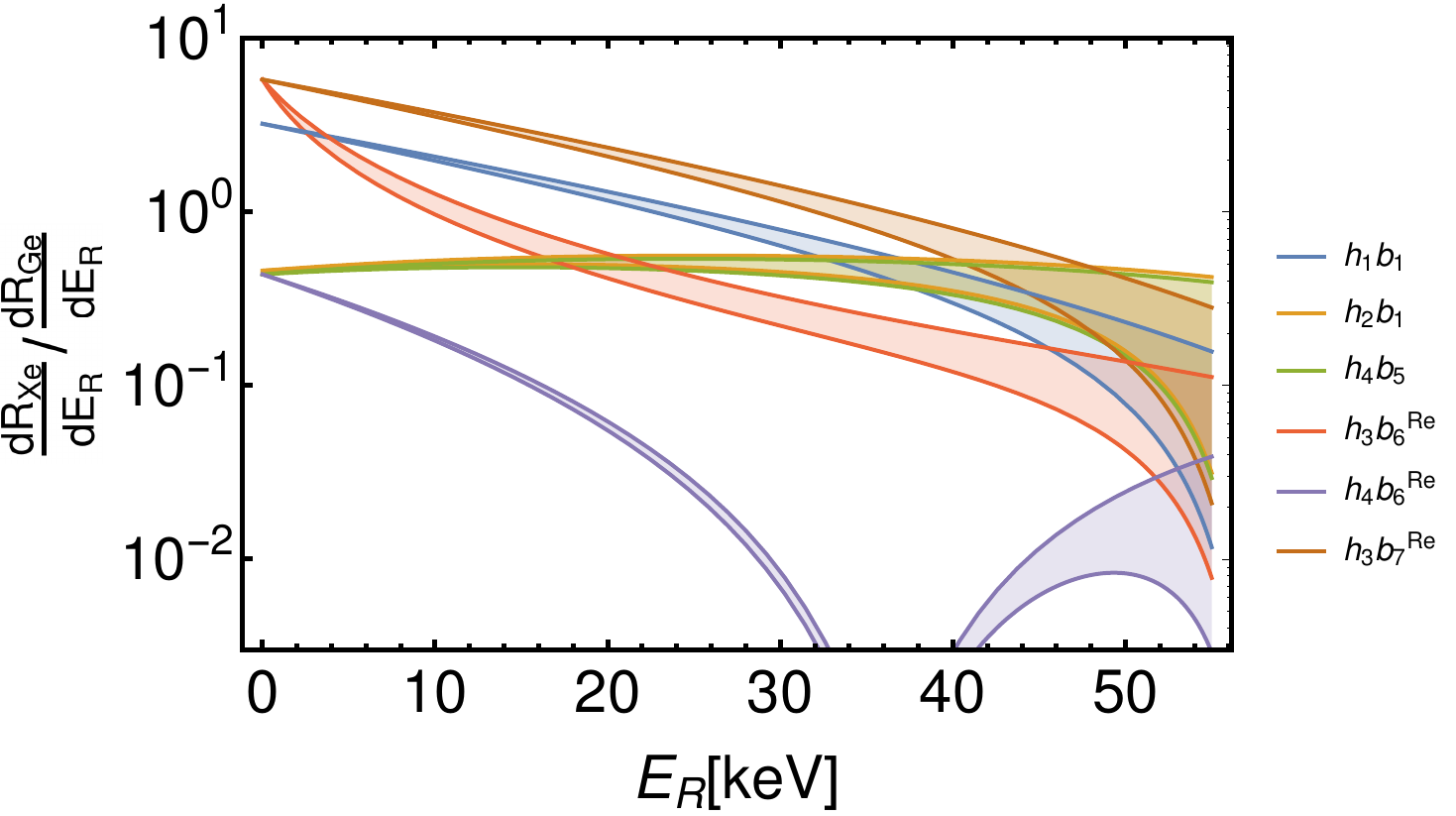}
}
\caption{Ratio of rates in xenon and germanium, illustrating the discriminating power of having multiple nuclear targets. For a 50GeV spin-$\half$ WIMP with uncharged mediator (left) and a 50GeV spin-1 WIMP with uncharged mediator (right), the shaded regions show the upper and lower bounds due to the astrophysical parameters}
\label{figRatios}
\end{figure}

To illustrate this discriminating power we plot the ratio of the rates in xenon and germanium in Fig.~\ref{figVectorI} and ~\ref{figRatios}. We choose to only present ratios for the uncharged mediator cases of spinor and vector WIMPs since the other cases produce trival results (all operators being spin independent). To estimate the effect astrophysical uncertainties will have on discriminating between operators, we plot the rate for a range of astrophysical parameters from $v_0=200$m/s, and $v_{esc}=500$m/s (lower) to $v_0=240$m/s and $v_{esc}=600$m/s (upper). The uncertainty in the dark matter density does not appear since we are considering the ratio of rates. Given the vastly different energy dependence of the ratio of rates of each scenario the astrophysical errors do not completely inhibit their identification.  Furthermore, operators $\mathcal{O}_9$ and $\mathcal{O}_{14}$, produced in scenarios $h_4 b_7^{\mathrm{Re}}$ and $h_4 b_6^{\mathrm{Re}}$ respectively, remain indistingushable when considering the ratio of rates. While it appears that in principle almost every operator is discernible, in practice isotopically impure targets and low statistics will further complicate the situation and provide limits on practical discrimination. \\

\section{Conclusion}

The analysis we have given here builds on previous analyses to provide, in generality, a roadmap to use event rates in direct dark matter detectors to constrain fundamental dark matter models. We have outlined the steps needed to go from fundamental Lagrangians, first to relativistic operators, then to non-relativistic operators, and finally to produce nuclear matrix elements.  In the process several significant facts have been elaborated.

\begin{itemize}

\item{Not all possible non-relativistic operators contributing to nuclear matrix elements in direct detection will arise from simple UV complete dark matter models.}

\item{Aside from scalar WIMPs each particular spin produces some non-relativistic operators that are unique to that spin.}

\item{Two non-relativistic operators, $\mathcal{O}_1$ and $\mathcal{O}_{10}$, are ubiquitous and arise for all WIMP spins we have explored.}

\item{In 5 scenarios, relativistic operators generate unique non-relativistic operators at leading order.}

\item{Two new non-relativistic operators not previously considered within the context of the full array of allowed nuclear responses} arise at low energies if spin-1 WIMP dark matter is allowed for.

\item{While the different operators that can contribute to event rates in detectors using specific elements or isotopes cannot be distinguished on the basis of their impact on the differential event rates in these detectors, they can produce radically different energy dependence for scattering off different nuclear targets.  Thus, a complementary use of different target materials will be necessary to reliably distinguish between different particle physics model possibilities for WIMP dark matter.}

\end{itemize}

While current detectors have only yielded upper limits, with new generations of larger detectors with greater energy resolution and lower thresholds coming online, the search for WIMP dark matter has never been so vibrant and promising. The tools we have provided here should help experimenters to probe the most useful parameter space, to interpret any non-zero signals in terms of constraints on fundamental models, and should allow theorists who build fundamental models to frame predictions in an accurate and simple way so that they might be directly compared with experiment.

\begin{acknowledgments}
We would like to thank Richard Lebed for useful discussions. J.B.D. thanks Dr. and Mrs. Sammie W. Cosper at the University of Louisiana at Lafayette, and the Louisiana Board of Regents for support. L.M.K, J.L.N and S.S acknowledge support from the DOE for this work under grant No.\,DE-SC0008016.  We thank the Australian National University for hospitality while part of this work was carried out.
\end{acknowledgments}
\begin{table}[h!]
\renewcommand{\arraystretch}{0.9}
\caption{List of scenarios with leading operators colored by which are distinguishable via the ratio $\frac{dR_{Xe}}{dE}/\frac{dR_{Ge}}{dE}$.}
\begin{tabular}{|c|c|C{.6cm}|C{.6cm}|C{.6cm}|C{.6cm}|C{.8cm}|C{.6cm}|C{.6cm}|C{.6cm}|C{.6cm}|C{.6cm}|C{.6cm}|C{.6cm}|C{.6cm}|C{.6cm}|C{.6cm}|C{.6cm}|C{.6cm}|C{.6cm}|}
\hline
& &\textcolor{pink}{$\mathcal{O}_{1}$} & $\mathcal{O}_2$ & $\mathcal{O}_3$ & \textcolor{green}{$\mathcal{O}_4$}& \textcolor{blue}{$q^2\mathcal{O}_4$} & \textcolor{red}{$\mathcal{O}_5$} &\textcolor{cyan}{$\mathcal{O}_6$} & $\mathcal{O}_7$ & \textcolor{purple}{$\mathcal{O}_8$} &\textcolor{orange}{$\mathcal{O}_9$} &\textcolor{blue}{$\mathcal{O}_{10}$} &\textcolor{magenta}{$\mathcal{O}_{11}$} &$\mathcal{O}_{12}$ &$\mathcal{O}_{13}$ & $\mathcal{O}_{14}$ & $\mathcal{O}_{15}$ & \textcolor{gray}{$\mathcal{O}_{17}$} & \textcolor{yellow}{$\mathcal{O}_{18}$} \\
\cmidrule[1pt]{2-20} 
& $(h_1,g_1)$      &\R&  &  &  &  &  &  &  &  &   &  &  &  &  &  &  &  &  \\
\cmidrule[.5pt]{2-20}
& $(h_2,g_1)$      &  &  &  &  &  &  &  &  &  &   &\B&  &  &  &  &  &  &  \\
\cmidrule[.5pt]{2-20}
& $(h_4,g_4)$      &  &  &  &  &  &  &  &  &  &   &\B&  &  &  &  &  &  &  \\
\cmidrule[.5pt]{2-20}
\rotatebox{90}{\rlap{Spin-$0$ WIMP}}
& $(y_1)$          &\R&  &  &  &  &  &  &  &  &   &\x&  &  &  &  &  &  &  \\
\cmidrule[.5pt]{2-20}
& $(y_2)$          &\R&  &  &  &  &  &  &  &  &   &\x&  &  &  &  &  &  &  \\
\cmidrule[.5pt]{2-20}
& $(y_1,y_2)$      &  &  &  &  &  &  &  &  &  &   &\B&  &  &  &  &  &  &  \\
\cmidrule[1pt]{1-20}
& $(h_1,\lambda_1)$&\R&  &  &  &  &  &  &  &  &   &  &  &  &  &  &  &  &  \\
\cmidrule[.5pt]{2-20}
&$(h_2,\lambda_1)$ &  &  &  &  &  &  &  &  &  &   &\B&  &  &  &  &  &  &  \\
\cmidrule[.5pt]{2-20}
&$(h_1,\lambda_2)$ &  &  &  &  &  &  &  &  &  &   &  &\M&  &  &  &  &  &  \\
\cmidrule[.5pt]{2-20}
&$(h_2,\lambda_2)$ &  &  &  &  &  &  &\C&  &  &   &  &  &  &  &  &  &  &  \\
\cmidrule[.5pt]{2-20}
&$(h_3,\lambda_3)$ &\R&  &  &  &  &  &  &  &  &   &  &  &  &  &  &  &  &  \\
\cmidrule[.5pt]{2-20}
&$(h_4,\lambda_3)$ &  &  &  &  &  &  &  &\x&  &\Or&  &  &  &  &  &  &  &  \\
\cmidrule[.5pt]{2-20}
&$(h_3,\lambda_4)$ &  &  &  &  &  &  &  &  &\Pu&\x&  &  &  &  &  &  &  &  \\
\cmidrule[.5pt]{2-20}
\rotatebox{90}{\rlap{Spin-$\half$ WIMP}}
&$(h_4,\lambda_4)$ &  &  &  &\G&  &  &  &  &  &   &  &  &  &  &  &  &  &  \\
\cmidrule[.5pt]{2-20}
&$(l_1)$           &\R&  &  &\x& &   &\x&  &  &   &  &  &  &  &  &  &  &  \\
\cmidrule[.5pt]{2-20}
&$(l_2)$           &\R&  &  &\x& &   &\x&  &  &   &  &  &  &  &  &  &  &  \\
\cmidrule[.5pt]{2-20}
&$(d_1)$           &\R&  &  &\x& &   &\x&  &  &   &  &  &  &  &  &  &  &  \\
\cmidrule[.5pt]{2-20}
&$(d_2)$           &\R&  &  &\x& &   &\x&  &  &   &  &  &  &  &  &  &  &  \\
\cmidrule[1pt]{1-20}
&$(h_1,b_1)$       &\R&  &  &  & &   &  &  &  &   &  &  &  &  &  &  &  &  \\
\cmidrule[.5pt]{2-20} 
&$(h_2,b_1)$       &  &  &  &  & &   &  &  &  &   &\B&  &  &  &  &  &  &  \\
\cmidrule[.5pt]{2-20}
&$(h_4,b_5)$       &  &  &  &  & &   &  &  &  &   &\B&  &  &  &  &  &  &  \\
\cmidrule[.5pt]{2-20}
&$(h_3,b_6)$       &  &  &  & &\x&\Red&\x&  &  &   &  &  &  &  &  &  &\g*&  \\
\cmidrule[.5pt]{2-20}
&$(h_4,b_6)$       &  &  &  &  & &   &  &  &  &\Or&  &  &  &  &  &  &  &\Y*\\
\cmidrule[.5pt]{2-20}
\rotatebox{90}{\rlap{Spin-$1$ WIMP}}
&$(h_3,b_7)$       &  &  &  &  &  &  &  &  &\Pu*&\x*&  &\M&  &  &  &  &  &  \\
\cmidrule[.5pt]{2-20}
&$(h_4,b_7)$       &  &  & &\G*&\rB&  &\x&  &  &   &  &  &  &  &\x&  &  &  \\
\cmidrule[.5pt]{2-20}
&$(y_3)$           &\R&  &  &\x& & &  &  &  &   &\x&\x&\x&  &  &  &  &\x\\
\cmidrule[.5pt]{2-20}
&$(y_4)$           &\R&  &  &\x& & &  &  &  &   &\x&\x&\x&  &  &  &  &\x\\
\cmidrule[.5pt]{2-20}
&$(y_3,y_4)$       &  &  &  &  & & &  &  &  &   &\x&\M&\x&  &  &  &  &\x\\
\hline
\end{tabular}
\label{tabFull}
\bfootnote{* indicates the purely imaginary scenario for that coupling}
\end{table}

\clearpage
\appendix

\section{Vector Dark Matter}
\label{appVec}
If the WIMP has spin $1$, we find two extra operators that haven't been considered previously. Specifically, the operators depend on the symmetric combination of polarization vectors, $S_{ij}=\half\left(\epsilon_i^\dagger\epsilon_j+\epsilon_j^\dagger\epsilon_i\right)$. This necessitates a modification to the WIMP response functions by first modifying the $\ell$ coefficients given in Eq.~\ref{eqnells}. Based on our non-relativistic reduction for vector dark matter, the Lagrangian for vector dark matter and the nucleus, interacting via an uncharged scalar or vector mediator can be written in general as:
\bea
\mathcal{L}_{vector}=c_1\mathcal{O}_1+c_4\mathcal{O}_4+c_5\mathcal{O}_5+c_8\mathcal{O}_8+c_9\mathcal{O}_9+c_{10}\mathcal{O}_{10}+c_{11}\mathcal{O}_{11}+c_{14}\mathcal{O}_{14}+c_{17}\mathcal{O}_{17}+c_{18}\mathcal{O}_{18}\nonumber\\
\eea
where we've defined $\mathcal{O}_{17}\equiv\frac{i\vec{q}}{m_N}\cdot\mathcal{S}\cdot\vec{v}_{\perp}$ and $\mathcal{O}_{18}\equiv\frac{i\vec{q}}{m_N}\cdot\mathcal{S}\cdot\vec{S}_{N}$ and the $c_i$'s are given in table \ref{tabCiVector}. To decompose these new operators we replace $\vec{v}^\perp$ with the target velocity and the internucleon velocities and sum over nucleons. $\mathcal{O}_{17}$ can then be put into the form
\bea
\mathcal{O}_{17}\rightarrow\frac{i\vec{q}}{m_N}.\mathcal{S}.\left[\vec{v}^\perp_T e^{-i\vec{q}.\vec{x}_i} - \sum_{i=1}^A\frac{1}{2M}\left(-\frac{1}{i}\overleftarrow{\nabla}_ie^{-i\vec{q}\cdot\vec{x}_i}+e^{-i\vec{q}\cdot\vec{x}_i}\frac{1}{i}\overrightarrow{\nabla}_i\right)_{int}\right].
\eea
$\mathcal{O}_{18}$ can be expanded as
\bea
\mathcal{O}_{18}\rightarrow\frac{1}{2}\frac{i\vec{q}}{m_N}\cdot\mathcal{S}\cdot\vec{\sigma}
\eea
Together, all the terms of $\mathcal{L}_{vector}$ give rise to the following $\ell$ factors from Eq.~\ref{eqnElls},
\bea
\ell_0^\tau&=&c_1^\tau+i\left( \frac{\vec{q}}{m_N}\times \vec{v}^\perp_T \right)\cdot \vec{S_\chi}c_5^\tau +(\vec{v}^{\perp}_T\cdot\vec{S}_\chi)c_8^\tau+i\left(\frac{\vec{q}}{m_N}\cdot\vec{S}_\chi\right)c_{11}^\tau+i\left(\frac{\vec{q}}{m_N}\cdot\mathcal{S}\cdot\vec{v}_{\perp}^T\right)c_{17}^\tau \nonumber\\
l_0^{A\tau}&=&-i\left(\frac{\vec{q}}{2m_N}\cdot\vec{S_\chi}\right)c_{14}^\tau \nonumber\\
\vec{l}_E^\tau&=&0 \\
\vec{l}_M^\tau&=&i\left(\frac{\vec{q}}{m_N}\times \vec{S}_\chi\right) c_5^\tau-\vec{S}_\chi c_8^\tau-i\left(\frac{\vec{q}}{m_N}\cdot\mathcal{S}\right)c_{17}^\tau \nonumber\\
\vec{l}_5^\tau&=&\frac{1}{2}\vec{S}_\chi c_4^\tau+i\left(\frac{\vec{q}}{m_N}\times\vec{S}_\chi\right)c_9^\tau+\frac{1}{2}\left(i\frac{\vec{q}}{m_N}\right)c_{10}^\tau+\frac{1}{2}\vec{v}^\perp_T\left(\frac{\vec{q}}{2m_N}\cdot\vec{S}_\chi\right)c_{14}^\tau+\frac{1}{2}\left(i\frac{\vec{q}}{m_N}\cdot\mathcal{S}\right)c_{18}^\tau \nonumber
\eea
Based on the $\ell$'s above, the coefficients of the various nuclear responses are found by squaring the amplitude and then summing over spins. To simplify calculations, we choose a convenient basis for polarization vectors, $\epsilon_{i}^s=\delta_{i}^s$. Recall that the spin can then be written as the anti-symmetric combination $iS_k=\epsilon_{ijk}\epsilon_i^\dagger\epsilon_j$. The WIMP responses unique to the vector case are then given by:
\bea
R_{M}^{\tau\tau'} &=& c_1^\tau c_1^{\tau'}+\frac{2}{3}\left(\frac{\vec{q}^2}{m_N^2}v_{T}^{\perp2}c_5^\tau c_5^{\tau'}+ v_{T}^{\perp2}c_8^\tau c_8^{\tau'}+\frac{q^2}{m_N^2}c_{11}^\tau c_{11}^{\tau'}+\frac{q^2v_{T}^{\perp2}}{4m_N^2}c_{17}^\tau c_{17}^{\tau'}\right) \nonumber\\
R_{\Phi^{''}}^{\tau\tau'}       &=& 0\nonumber\\
R_{\Phi^{''}M}^{\tau\tau'}      &=& 0\nonumber\\
R_{\tilde{\Phi}^{'}}^{\tau\tau'}&=& 0\nonumber\\
R_{\Sigma^{''}}^{\tau\tau'}     &=& \frac{1}{6}c_4^\tau c_4^{\tau'}+\frac{q^2}{4m_N^2}c_{10}^\tau c_{10}^{\tau'}+\frac{q^2}{12m_N^2}c_{18}^\tau c_{18}^{\tau'} \nonumber\\
R_{\Sigma^{'}}^{\tau\tau'}      &=& \frac{1}{6}c_4^\tau c_4^{\tau'}+\frac{q^2}{6m_N^2}c_{9}^\tau c_{9}^{\tau'}+\frac{q^2v_{T}^{\perp2}}{2m_N^2}c_{14}^\tau c_{14}^{\tau'}+\frac{q^2}{24m_N^2}c_{18}^\tau c_{18}^{\tau'} \nonumber\\
R_{\Delta}^{\tau\tau'}&=&\frac{2}{3}\left(\frac{\vec{q}^2}{m_N^2}c_5^\tau c_5^{\tau'}+c_8^\tau c_8^{\tau'}\right)+\frac{q^2}{6m_N^2}c_{17}^\tau c_{17}^{\tau'} \nonumber\\
R_{\Delta\Sigma^{'}}^{\tau\tau'}&=&\frac{2}{3}\left(c_5^\tau c_4^{\tau'}-c_8^\tau c_9^{\tau'}\right).\\ \nonumber 
\eea

\section{Non-relativistic Reduction}
\label{appNRR}

We find effective relativistic interaction Lagrangians by integrating out heavy mediators. We only keep the leading order interactions (suppressed by $m$ or $m^2$). To the right of each operator is their non-relativistic reduction expressed in terms of the operators in table~\ref{tabHaxtonOp} with the coefficient derived from the Lagrangian parameters along with the relevant nucleon form factor. As multiple operators can have the same non-relativistic limit, it is important to include the nucleon form factor at the relativistic level.  If this is not performed, erroneous cancellations can occur. 

For free spinors we use the Bjorken and Drell normalization and $\gamma$ matrix conventions. In the non-relativistic limit we make the following replacements:
\bea
S &\rightarrow& \frac{1_{S}}{\sqrt{m_S}} \nonumber\\
X_{\mu}  &\rightarrow& \frac{\epsilon_{\mu}^s}{\sqrt{m_X}} \nonumber\\
\chi     &\rightarrow& \sqrt{\frac{E+m_\chi}{2m_\chi}} \begin{pmatrix} \xi\\ \frac{\vec{\sigma}\cdot\vec{p}}{E+m_\chi}\xi   \end{pmatrix} \\ \nonumber
\eea
where $s=1,2,3$ are the different polarization states of the vector.  $\xi=(1\ 0)^T$ is the left handed Weyl spinor. The following Fierz transformation and gamma matrix identites were useful in the charged mediator cases, (a sign difference was found in the final identity when compared with \cite{Agrawal:2010}):
\bea
(\bar{q}\chi)(\bar{\chi}q) =&& -\frac{1}{4}\left[ \bar{q}q\bar{\chi}\chi + \bar{q}\gamma^{\mu}q\bar{\chi}\gamma_{\mu}\chi + \frac{1}{2}\bar{q}\sigma^{\mu\nu}q\bar{\chi}\sigma_{\mu\nu}\chi-\bar{q}\gamma^{\mu}\gamma^5q\bar{\chi}\gamma_{\mu}\gamma^5\chi + \bar{q}\gamma^5q\bar{\chi}\gamma^5\chi\right] \nonumber\\
(\bar{q}\gamma^5\chi)(\bar{\chi}\gamma^5q) = &&-\frac{1}{4}\left[\bar{q}q\bar{\chi}\chi + \bar{q}\gamma^5q\bar{\chi}\gamma^5\chi -\bar{q}\gamma^{\mu}q\bar{\chi}\gamma_{\mu}\chi+ \bar{q}\gamma^{\mu}\gamma^5q\bar{\chi}\gamma_{\mu}\gamma^5\chi + \frac{1}{2}\bar{q}\sigma^{\mu\nu}q\bar{\chi}\sigma_{\mu\nu}\chi\right] \nonumber\\
(\bar{q}\chi)(\bar{\chi}\gamma^5q) = && -\frac{1}{4}\left[\bar{q}q\bar{\chi}\gamma^5\chi + \bar{q}\gamma^5q\bar{\chi}\chi -\bar{q}\gamma^{\mu}q\bar{\chi}\gamma_{\mu}\gamma^5\chi+ \bar{q}\gamma^{\mu}\gamma^5q\bar{\chi}\gamma_{\mu}\chi +i\epsilon_{\mu\nu\alpha\beta}\bar{q}\sigma^{\mu\nu}q\bar{\chi}\sigma^{\alpha\beta}\chi\right] \nonumber\\
(\bar{q}\gamma_{\mu}\chi)(\bar{\chi}\gamma^{\mu}q) = && -\left[\bar{q}q\bar{\chi}\chi - \bar{q}\gamma^5q\bar{\chi}\gamma^5\chi -\frac{1}{2}\bar{q}\gamma^{ \mu}q\bar{\chi}\gamma_{ \mu}\chi - \frac{1}{2}\bar{q}\gamma^{ \mu}\gamma^5q\bar{\chi}\gamma_{ \mu}\gamma^5\chi\right]\nonumber\\
(\bar{q}\gamma_{\mu}\gamma^5\chi)(\bar{\chi}\gamma^{\mu}\gamma^5q) = && -\left[-\bar{q}q\bar{\chi}\chi + \bar{q}\gamma^5q\bar{\chi}\gamma^5\chi-\frac{1}{2}\bar{q}\gamma^{ \mu}q\bar{\chi}\gamma_{ \mu}\chi -\frac{1}{2}\bar{q}\gamma^{ \mu}\gamma^5q\bar{\chi}\gamma_{ \mu}\gamma^5\chi\right]\nonumber\\
(\bar{q}\gamma_{\mu}\chi)(\bar{\chi}\gamma^{\mu}\gamma^5q) = && -\left[\bar{q}q\bar{\chi}\gamma^5\chi - \bar{q}\gamma^5q\bar{\chi}\chi + \frac{1}{2}\bar{q}\gamma^{ \mu}q\bar{\chi}\gamma_{\mu}\gamma^5\chi +\frac{1}{2}\bar{q}\gamma^{ \mu}\gamma^5q\bar{\chi}\gamma_{ \mu}\chi\right]
\eea
\bea
 \sigma^{\mu\nu}\gamma^5 = \frac{i}{2}\epsilon^{\mu\nu\rho\sigma}\sigma_{\rho\sigma}
\eea
All of the following operators are collected in terms of the coefficients of the NR operators, $c_i$, in tables \ref{tabCiScalar},\ref{tabCiSpinor} and \ref{tabCiVector}. 

\begin{table}[ht]
\caption{Non-relativistic reduction of operators for a spin-$0$ WIMP}
	\begin{tabular}{lcl}
\hline
Scalar Mediator \\
\hline
		$( S^\dag S)(\bar{q}q)$ &$\longrightarrow$& $\left(\frac{h_1^Ng_{1}}{m_{\phi}^2}\right)\mathcal{O}_1$\\
		$( S^\dag S)(\bar{q}\gamma^5q)$ &$\longrightarrow$& $\left(\frac{h_2^Ng_{1}}{m_{\phi}^2}\right)\mathcal{O}_{10}$\\
\hline
Vector Mediator \\
\hline			
	$i(S^{\dag}\partial_{\mu}S-\partial_{\mu}S^{\dag}S)(\bar{q}\gamma^{\mu}q)$ &$\longrightarrow$&$0$\\
	$i(S^{\dag}\partial_{\mu}S-\partial_{\mu}S^{\dag}S)(\bar{q}\gamma^{\mu}\gamma^5q)$ &$\longrightarrow$&$ \left(\frac{2ig_4h_4^N}{m_G^2}\frac{m_N}{m_S}\right)\mathcal{O}_{10}$\\
\hline
Charged Spinor Mediator \\
\hline
 $(S^{\dag}S)(\bar{q}q)$         &$\longrightarrow$ & $\frac{y_1^\dag y_1-y_2^\dag y_2}{m_Qm_S}f_{T}^N\mathcal{O}_1$\\
 $(S^{\dag}S)(\bar{q}\gamma^5q)$ &$\longrightarrow$ & $i\frac{y_2^\dag y_1-y_1^\dag y_2}{m_Qm_S}\tilde\Delta^N\mathcal{O}_{10}$\\
	\end{tabular}
\end{table}

\begin{table}[ht]
\label{NRspinorNeutral}
\caption{Operators for a spin-$\half$ WIMP via a neutral mediator}
\begin{tabular}{lcl}	
\hline
Scalar Mediator \\
\hline
	 $\bar{\chi}\chi\bar{q}q$ & $\longrightarrow$ & $\left( \frac{h_1^N \lambda_1}{m_\phi^2}\right)\mathcal{O}_{1}$\\
	 $\bar{\chi}\chi\bar{q}\gamma^5q$ & $\longrightarrow$ & $\left( \frac{h_2^N \lambda_1}{m_\phi^2}\right)\mathcal{O}_{10}$ \\
	 $\bar{\chi}\gamma^5\chi\bar{q}q$ & $\longrightarrow$ & $\left( -\frac{h_1^N \lambda_2m_N}{m_\phi^2 m_\chi}\right)\mathcal{O}_{11}$ \\
	 $\bar{\chi}\gamma^5\chi\bar{q}\gamma^5q$ & $\longrightarrow$ & $\left( \frac{h_2^N \lambda_2m_N}{m_\phi^2m_\chi}\right)\mathcal{O}_{6}$ \\
\hline
Vector Mediator \\
\hline	
	$\bar{\chi}\gamma^\mu\chi\bar{q}\gamma_\mu q$             &$\longrightarrow$&$\left(-\frac{h_3^N\lambda_3}{m_G^2}\right)\mathcal{O}_1$\\
   $\bar{\chi}\gamma^\mu\chi\bar{q}\gamma_\mu \gamma^5q$     &$\longrightarrow$&$\left(-\frac{2h_4^N\lambda_3}{m_G^2}\right)\left(-\mathcal{O}_7+\frac{m_N}{m_\chi}\mathcal{O}_9\right)$\\
	$\bar{\chi}\gamma^\mu\gamma^5\chi\bar{q}\gamma_\mu q$     &$\longrightarrow$&$\left(-\frac{2h_3^N\lambda_4}{m_G^2}\right)\left(\mathcal{O}_8+\mathcal{O}_9\right)$\\
	$\bar{\chi}\gamma^\mu\gamma^5\chi\bar{q}\gamma_\mu\gamma^5q$            & $\longrightarrow$ & $\left( \frac{4h_4^N \lambda_4}{m_G^2}\right)\mathcal{O}_4$\\
		\end{tabular}
\end{table}

\begin{table}[ht]	
\caption{Non-relativistic reduction of operators for a spin-$\half$ WIMP via a charged mediator (after using Fierz identities)}
\label{NRspinorCharged}
\begin{tabular}{lcl}
\hline
Charged Scalar Mediator & & \\
\hline
	$\bar{\chi}\chi\bar{q}q$ &                 $\longrightarrow$  & $ \frac{l_2^\dag l_2 - l_1^\dag l_1}{4 m_\Phi^2}f_{Tq}^N\mathcal{O}_{1}$\\
	$\bar{\chi}\chi\bar{q}\gamma^5q$ &         $\longrightarrow$ & $i\frac{l_1^\dag l_2 - l_2^\dag l_1}{4 m_\Phi^2}\Delta\tilde{q}^N\mathcal{O}_{10}$ \\ 
	$\bar{\chi}\gamma^5\chi\bar{q}q$ &         $\longrightarrow$ & $i\frac{l_2^\dag l_1 - l_1^\dag l_2}{4 m_\Phi^2}\frac{m_N}{m_\chi}f_{Tq}^N\mathcal{O}_{11}$ \\
	$\bar{\chi}\gamma^5\chi\bar{q}\gamma^5q$ & $\longrightarrow$  & $ \frac{l_1^\dag l_1 - l_2^\dag l_2}{4 m_\Phi^2}\frac{m_N}{m_\chi}\Delta\tilde{q}^N\mathcal{O}_{6}$ \\
	$\bar{\chi}\gamma^\mu\chi\bar{q}\gamma_\mu q$            & $\longrightarrow$ & $-\frac{l_1^\dag l_1 + l_2^\dag l_2}{4 m_\Phi^2}\mathcal{N}_q^N\mathcal{O}_1$\\
   $\bar{\chi}\gamma^\mu\gamma^5\chi\bar{q}\gamma_\mu q$    & $\longrightarrow$ & $\frac{l_1^\dag l_2 + l_2^\dag l_1}{2 m_\Phi^2}\mathcal{N}_q^N(\mathcal{O}_8+\mathcal{O}_9)$\\
   $\bar{\chi}\gamma^\mu\chi\bar{q}\gamma_\mu\gamma^5q$     & $\longrightarrow$ & $\frac{l_1^\dag l_2 + l_2^\dag l_1}{2 m_\Phi^2}\Delta_{q}^N(\mathcal{O}_7-\frac{m_N}{m_\chi}\mathcal{O}_9)$\\
   $\bar{\chi}\gamma^\mu\gamma^5\chi\bar{q}\gamma_\mu\gamma^5q$  & $\longrightarrow$ & $-\frac{l_1^\dag l_1 + l_2^\dag l_2}{m_\Phi^2}\Delta_{q}^N\mathcal{O}_4$\\
	$\bar{\chi}\sigma^{\mu\nu}\chi\bar{q}\sigma_{\mu\nu} q$   		 & $\longrightarrow$ & $ \frac{l_2^\dag l_2 - l_1^\dag l_1}{m_\Phi^2}\delta_{q}^N\mathcal{O}_4$\\
	$\epsilon_{\mu\nu\alpha\beta}\bar{\chi}\sigma^{\mu\nu}\chi\bar{q}\sigma^{\alpha\beta} q$ 	&$\longrightarrow$ & $\frac{l_2^\dag l_1 - l_1^\dag l_2}{m_\Phi^2}\delta_{q}^N(i\mathcal{O}_{10}-i\frac{m_N}{m_\chi}\mathcal{O}_{11}+4\mathcal{O}_{12})$\\
\hline
Charged Vector Mediator & & \\
\hline
   $\bar{\chi}\chi\bar{q}q$ &         $\longrightarrow$ & $ \frac{d_2^\dag d_2 - d_1^\dag d_1}{4m_V^2}f_{Tq}^N\mathcal{O}_{1}$\\
	$\bar{\chi}\chi\bar{q}\gamma^5q$ & $\longrightarrow$ & $i\frac{d_2^\dag d_1 - d_1^\dag d_2}{4m_V^2}\Delta\tilde{q}^N\mathcal{O}_{10}$ \\
	$\bar{\chi}\gamma^5\chi\bar{q}q$ & $\longrightarrow$ & $i\frac{d_2^\dag d_1 - d_1^\dag d_2}{4m_V^2}\frac{m_N}{m_\chi}f_{Tq}^N\mathcal{O}_{11}$ \\
	$\bar{\chi}\gamma^5\chi\bar{q}\gamma^5q$ & $\longrightarrow$ & $\frac{d_2^\dag d_2 - d_1^\dag d_1}{4m_V^2}\frac{m_N}{m_\chi}\Delta\tilde{q}^N\mathcal{O}_{6}$ \\
	$\bar{\chi}\gamma^\mu\chi\bar{q}\gamma_\mu q$               &$\longrightarrow$ & $\frac{d_2^\dag d_2 + d_1^\dag d_1}{8m_V^2}\mathcal{N}_q^N\mathcal{O}_1$\\
   $\bar{\chi}\gamma^\mu\gamma^5\chi\bar{q}\gamma_\mu q$       &$\longrightarrow$ & $-\frac{d_2^\dag d_1 + d_1^\dag d_2}{4 m_V^2}\mathcal{N}_q^N(\mathcal{O}_8+\mathcal{O}_9)$\\
   $\bar{\chi}\gamma^\mu\chi\bar{q}\gamma_\mu\gamma^5 q$       &$\longrightarrow$ & $\frac{d_2^\dag d_1 + d_1^\dag d_2}{4m_V^2}\Delta_{q}^N(\mathcal{O}_7-\frac{m_N}{m_\chi}\mathcal{O}_9)$\\
   $\bar{\chi}\gamma^\mu\gamma^5\chi\bar{q}\gamma_\mu\gamma^5 q$&$\longrightarrow$& $-\frac{d_2^\dag d_2 + d_1^\dag d_1}{2m_V^2}\Delta_{q}^N\mathcal{O}_4$\\
 \end{tabular}
\end{table}

\clearpage
\begin{table}[ht]
\caption{Non-relativistic reduction of operators for a spin-$1$ WIMP}
\begin{tabular}{lcl}
\hline 
Scalar Mediator & & \\
\hline	
 $X_{\mu}^{\dag}X^{\mu}\bar{q}q$                   &       $\longrightarrow$ & $\left( \frac{b_1h_1^N}{m_{\phi}^2}\right)\mathcal{O}_1$\\
 $X_{\mu}^{\dag}X^{\mu}\bar{q}\gamma^5q$           &       $\longrightarrow$ & $\left( \frac{b_1h_2^N}{m_{\phi}^2}\right)\mathcal{O}_{10}$\\
\hline 
Vector Mediator & & \\
\hline
 $(X_{\nu}^{\dagger}\partial_{\mu}X^{\nu}-\partial_{\mu}X_{\nu}^{\dagger}X^{\nu})(\bar{q}\gamma^{\mu}q)$     &    $\longrightarrow$ & $0$\\
 $(X_{\nu}^{\dagger}\partial_{\mu}X^{\nu}-\partial_{\mu}X_{\nu}^{\dagger}X^{\nu})(\bar{q}\gamma^{\mu}\gamma^{5}q)$  &  $\longrightarrow$ & $\left(\frac{-3b_5h_4^N}{m_G^2}\frac{m_N}{m_X}\right)\mathcal{O}_{10}$\\	
$\partial_{\nu}(X^{\nu\dagger}X_{\mu}+X^\dagger_\mu X^{\nu})(\bar{q}\gamma^{\mu}q)$   &    $\longrightarrow$ & $\left( \frac{\mathrm{Re}(b_6)h_3^N}{m_G^2}\frac{m_N}{m_X}\right)(\mathcal{O}_{5}+\mathcal{O}_6-\frac{q^2}{m_N^2}\mathcal{O}_4)$\\	
 $\partial_{\nu}(X^{\nu\dagger}X_{\mu}+X^\dagger_\mu X^\nu)(\bar{q}\gamma^{\mu}\gamma^{5}q)$      &   $\longrightarrow$ & $\left( -\frac{2\mathrm{Re}(b_6)h_4^N}{m_G^2}\frac{m_N}{m_X}\right)\mathcal{O}_{9}$\\
 $\partial_{\nu}(X^{\nu\dagger}X_{\mu}-X^\dagger_\mu X^{\nu})(\bar{q}\gamma^{\mu}q)$   &    $\longrightarrow$ & $\left( -\frac{4\mathrm{Im}(b_6)h_3^N}{m_G^2}\frac{m_N}{m_X}\right)\mathcal{O}_{17}$\\	
 $\partial_{\nu}(X^{\nu\dagger}X_{\mu}-X^\dagger_\mu X^\nu)(\bar{q}\gamma^{\mu}\gamma^{5}q)$      &   $\longrightarrow$ & $\left( \frac{4\mathrm{Im}(b_6)h_4^N}{m_G^2}\frac{m_N}{m_X}\right)\mathcal{O}_{18}$\\
 $\epsilon_{\mu\nu\rho\sigma}\left(X^{\nu\dag}\partial^{\rho}X^\sigma+X^\nu\partial^\rho X^{\sigma\dag}\right)(\bar{q}\gamma^{\mu}q)$    &    $\longrightarrow$ & $\left( \frac{\mathrm{Re}(b_7)h_3^N}{m_G^2}\frac{m_N}{m_X}\right)\mathcal{O}_{11}$\\
 $\epsilon_{\mu\nu\rho\sigma}\left(X^{\nu\dag}\partial^{\rho}X^\sigma+X^\nu\partial^\rho X^{\sigma\dag}\right)(\bar{q}\gamma^{\mu}\gamma^5 q)$    &    $\longrightarrow$ & $\left(\frac{\mathrm{Re}(b_7)h^N_4}{m_G^2}\frac{m_N}{m_X}\right)(i\frac{q^2}{m_Xm_N}\mathcal{O}_4-i\frac{m_N}{m_X}\mathcal{O}_6-2\mathcal{O}_{14})$\\
 $\epsilon_{\mu\nu\rho\sigma}\left(X^{\nu\dag}\partial^{\rho}X^\sigma-X^\nu\partial^\rho X^{\sigma\dag}\right)(\bar{q}\gamma^{\mu}q)$  &   $\longrightarrow$ & $\left( \frac{2\mathrm{Im}(b_7)h_3^N}{m_G^2}\right)(\mathcal{O}_8+\mathcal{O}_9)$\\
 $\epsilon_{\mu\nu\rho\sigma}\left(X^{\nu\dag}\partial^{\rho}X^\sigma-X^\nu\partial^\rho X^{\sigma\dag}\right)(\bar{q}\gamma^{\mu}\gamma^5q)$  &   $\longrightarrow$ & $\left( \frac{4\mathrm{Im}(b_7)h_4^N}{m_G^2}\right)\mathcal{O}_4$\\

\hline
Charged Spinor Mediator \\
\hline
	 $(X_{\mu}^\dag X_{\nu})(\bar{q}\gamma^{\mu}\gamma^{\nu}q)$        &    $\longrightarrow$ & $\left(\frac{y_3^\dag y_3-y_4^\dag y_4}{m_Q m_X}\right)\left(f_{Tq}^N\mathcal{O}_1+2\delta_{q}^N\mathcal{O}_4\right)$\\
	 $(X_{\mu}^\dag X_{\nu})(\bar{q}\gamma^{\mu}\gamma^{\nu}\gamma^5q)$       &     $\longrightarrow$ & $\left(\frac{y_4^\dag y_3-y_3^\dag y_4}{m_Q m_X}\right)(i\Delta^N_{\tilde q}\mathcal{O}_{10}+i\delta_{q}^N\mathcal{O}_{11}-2i\delta_{q}^N \mathcal{O}_{12}-2i\delta_{q}^N\mathcal{O}_{18})$\\
		\end{tabular}
\end{table}

\section{Quarks to Nucleons}
\label{appQ2N}
To go from the fundamental interactions of WIMPs with quarks to scattering from point-like nucleons, one must evaluate the quark (parton) bilinears in the nucleons. For a full discussion see the appendix of \cite{Agrawal:2010} and \cite{Dienes:2013}. We write the nucleon couplings in terms of the quark couplings times a form factor (in the limit of zero momentum transfer):
\begin{table}[h!]
\centering
\begin{tabular}{lrl}
 $\Bra{N_o}m_q \qbar q        \Ket{N_i}$ &     $\longrightarrow$&  $f^N_{Tq} \Nbar N$\\
 $\Bra{N_o}\qbar \gamma^5 q   \Ket{N_i}$ &     $\longrightarrow$&  $\Delta\tilde{q}^N \Nbar \gamma^5 N$\\
 $\Bra{N_o}\qbar \gamma^\mu q \Ket{N_i}$ &     $\longrightarrow$&  $\mathcal{N}^N_q \Nbar \gamma^\mu N$\\
 $\Bra{N_o}\qbar\gamma^\mu\gamma^5q\Ket{N_i}$& $\longrightarrow$&  $\Delta^N_q\Nbar\gamma^\mu\gamma^5N$\\
 $\Bra{N_o}\qbar\sigma^{\mu\nu} q\Ket{N_i}$&   $\longrightarrow$&  $\delta^N_q\Nbar\sigma^{\mu\nu} N$\\
\end{tabular}
\end{table}
The scalar bilinear for light quarks can be evaluated from
\be
\Bra{N} m_q \qbar q \Ket{N} = m_N f^N_{Tq}
\ee
while for the heavy quarks 
\be
\Bra{N} m_q \qbar q \Ket{N} = \frac{2}{27} m_N F^N_{TG} = \frac{2}{27} m_N \left( 1 - \sum_{q=u,d,s}f^N_{Tq}\right).
\ee
Summing over all the quarks one finds
\be
h_1^N = \sum_{q=u,d,s} h_1^q \frac{m_N}{m_q} f^N_{Tq} + \frac{2}{27} f^N_{TG} \sum_{q=c,b,t} h_1^q \frac{m_N}{m_q}
\ee

The psuedo-scalar bilinear was recently revisited in \cite{Dienes:2013}:
\be
h_2^N = \sum_{q=u,d,s} h_2^q \Delta\tilde{q}^N - \Delta \tilde{G}^N\sum_{q=c,b,t} \frac{h_2^q }{m_q}
\ee

The vector bilinear essentially gives the number operator:
\be
h_3^N  = \left\{
  \begin{array}{lr}
    2 h_3^u + h_3^d & \,\, N=p\\
    h_3^u + 2 h_3^d & \,\, N=u\\
  \end{array}
\right.
\ee

The psuedo-vector bilinear counts the contributions of spin to the nucleon (note that sometimes this coupling has a $G_F$ factored out to make it dimensionless) 
\be
h_4^N = \sum_{q=u,d,s} h_4^q \Delta^N_{q}
\ee

Throughout this paper the following values are used (it should be noted that there are large uncertainties in these values)~\cite{Agrawal:2010,Dienes:2013}:
\bea
f^n_{Tu} = & 0.014  \,\,\,\,\,  &  f^p_{Tu}  = 0.02 \nonumber\\
f^n_{Td} = & 0.036 	\,\,\,\,\,	  &  f^p_{Td}   = 0.026\nonumber\\
f^n_{Ts} = & 0.118  \,\,\,\,\,		  &  f^p_{Ts} = 0.118\nonumber\\
\Delta^n_{u} = & -0.427  \,\,\,\,\, & \Delta^p_{u} = 0.842 \nonumber\\
\Delta^n_{d} = & 0.842  \,\,\,\,\,  & \Delta^p_{d} = -0.427 \nonumber\\
\Delta^n_{s} = & -0.085 \,\,\,\,\,  & \Delta^p_{s} = -0.085 \nonumber\\
\Delta\tilde{u}^n = & -108.03  \,\,\,\,\, & \Delta\tilde{u}^p = 110.55 \nonumber\\
\Delta\tilde{d}^n = & 108.60  \,\,\,\,\,  & \Delta\tilde{d}^p = -107.17 \nonumber\\
\Delta\tilde{s}^n = & -0.57 \,\,\,\,\,  & \Delta\tilde{s}^p = -3.37 \nonumber\\
\Delta\tilde{G}^n = & 35.7 \mathrm{ MeV}\,\,\,\,\,& \Delta\tilde{G}^p = 395.2\mathrm{ MeV} \nonumber\\
\eea
Assuming a universal coupling of the mediators to all quarks, the nucleon level couplings can then be written as,
\bea
 h_1^N=& f_{T}^N h_1 \nonumber\\
 h_2^N=& \tilde\Delta^N h_2 \nonumber\\
 h_3^N=& \mathcal{N}^N h_3     \nonumber\\
 h_4^N=& \Delta^N h_4  \nonumber\\
\eea
where we have defined,
\bea
 f_{T}^n=& 11.93 \,\,\,\,\,& f_{T}^p= 12.31 \nonumber\\
 \tilde\Delta^n=& -0.07  \,\,\,\,\,& \tilde\Delta^p= -0.28 \nonumber\\
 \mathcal{N}^n=& 3      \,\,\,\,\,& \mathcal{N}^p= 3    \nonumber\\
 \Delta^n=& 0.33   \,\,\,\,\,& \Delta^p= 0.33 \nonumber\\
 \delta^n=& 0.564   \,\,\,\,\,& \delta^p= 0.564 \nonumber\\.
\eea
This introduces a small amount of isospin violation, and it is known that relaxing the assumption of universal couplings to quarks can lead to interesting isospin violating effects~\cite{Dienes:2013,Feng:2013vod}.

\end{document}